%% file: Plancherel_growth_modified.tex
\RequirePackage{ifpdf}
\documentclass[final,3p,12pt,times]{elsarticle}
\pdfoutput=1

\usepackage{amssymb,amsmath,xcolor,float}
\usepackage{amsthm}
\usepackage{lineno}
\usepackage{hyperref}
\usepackage{cleveref}
\numberwithin{equation}{section}

\usepackage{showkeys}
\usepackage[inline]{showlabels}
\usepackage{tabularx}
\usepackage{youngtab}
\usepackage{young}
\usepackage{wrapfig}
\usepackage{graphicx}

\usepackage{float}
\usepackage[font=small,labelfont=bf]{caption}
\usepackage{mathrsfs}
\usepackage[parfill]{parskip}
\usepackage{subcaption}

\input{gdefn.tex}

\makeatletter
\def\ps@pprintTitle{%
	\let\@oddhead\@empty
	\let\@evenhead\@empty
	\def\@oddfoot{}%
	\let\@evenfoot\@oddfoot}
\makeatother

\biboptions{sort&compress}
\begin{document}
	
	\begin{frontmatter}
		
		
		
		\title{Quantum Mechanics of Plancherel Growth}

		
		\author[1,2]{Arghya Chattopadhyay}
		\ead{arghya.chattopadhyay@gmail.com}
		\author[3]{Suvankar Dutta}
		\ead{suvankar@iiserb.ac.in}
		\author[3,4]{Debangshu Mukherjee}
		\ead{debangshu0@gmail.com}
		\author[3]{Neetu}
		\ead{neetuj@iiserb.ac.in}
		
		
		\address[1]{Institute of Mathematical Sciences, Homi Bhaba National Institute (HBNI),\\ IV Cross Road, Taramani, Chennai 600113, India}
		\address[2]{National Institute for Theoretical Physics,\\School of Physics and Mandelstam Institute for Theoretical Physics,\\University of the Witwatersrand, Wits, 2050, South Africa.}
		\address[3]{Department of Physics, Indian Institute of Science Education and Research  Bhopal,\\ Bhopal bypass, Bhopal 462066, India}
		\address[4]{School of Physics, Indian Institute of Science Education and Research Thiruvananthapuram,\\ Vithura 695551, Kerala, India.}

\begin{abstract}
Growth of Young diagrams, equipped with Plancherel measure, follows the automodel equation of Kerov. Using the technology of unitary matrix model we show that such growth process is exactly same as the growth of gap-less phase in Gross-Witten and Wadia (GWW) model. The \emph{limit shape} of asymptotic Young diagrams corresponds to GWW transition point. Our analysis also offers an alternate proof of \emph{limit shape} theorem of Vershik-Kerov and Logan-Shepp. Using the connection between unitary matrix model and free Fermi droplet description, we map the Young diagrams in automodel class to different shapes of two dimensional phase space droplets. Quantising these droplets we further set up a correspondence between automodel diagrams and coherent states in the Hilbert space. Thus growth of Young diagrams are mapped to evolution of coherent states in the Hilbert space. Gaussian fluctuations of large \emph{N} Young diagrams are also mapped to quantum (large \emph{N}) fluctuations of the coherent states.

\end{abstract}
		

\end{frontmatter}

\tableofcontents


\section{Introduction}
\label{sec:intro}

The theory of random matrices has wide applications both in physics and mathematics. A particular class of random matrix model namely the unitary matrix model (UMM) plays a pivotal role in understanding the various thermodynamic properties, phase structure  and dynamics of gauge theories in diverse dimensions. In this paper we explore yet another interesting application of UMM in the field of asymptotic growth of Young diagrams in representation theory. Using the techniques of unitary matrix models we provide a Hilbert space description of this growth of Young diagrams - the \emph{Plancherel growth}.

\begin{figure}[h]
	\begin{center}
		\includegraphics[width=10.5cm,height=8cm]{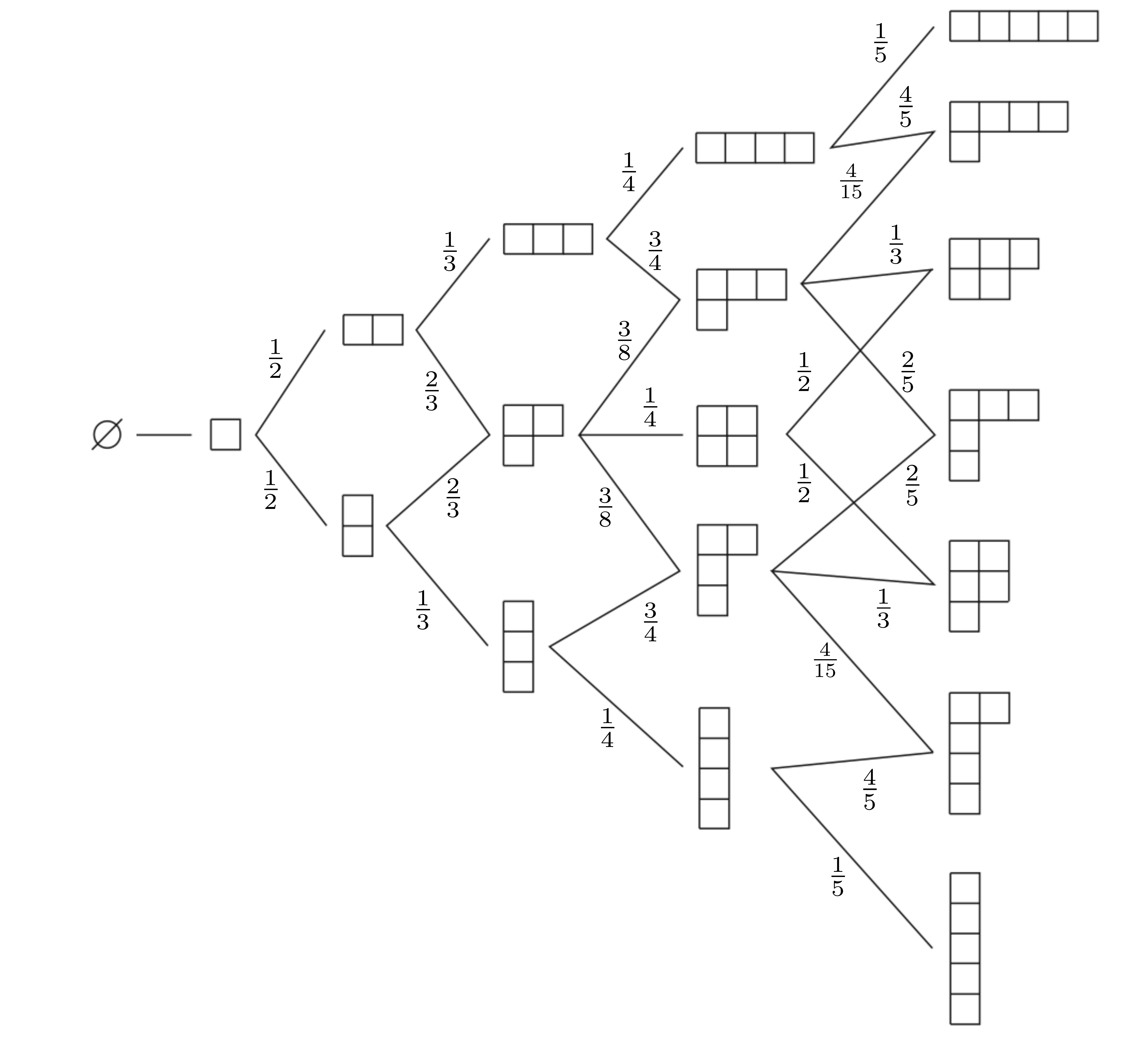}
		\caption{\footnotesize{Growth of Young diagrams.}}
		\label{fig:younggrowth}
	\end{center}
\end{figure}

Young diagrams provide a convenient diagrammatic way to describe the representations of symmetric group and general linear groups. There are different notations in literature to depict a Young diagram. In this paper we follow the ``English notation". In this notation, boxes are arranged in horizontal rows with the condition that number of boxes in a row is always less than or equal to that in the row above. In general as one goes to higher and higher dimensional representations the number of boxes in Young diagram increases. For our purpose we classify the Young diagrams in terms of total number of boxes. Let us denote the set of all Young diagrams with $k$ boxes by $\cY_k$. All the diagrams in $\cY_{k+1}$ can be obtained by adding one box to each diagram in $\cY_k$ in all possible allowed ways. See figure \ref{fig:younggrowth}.
Such a process is called \emph{growth process} of Young diagrams. One can construct all possible Young diagrams at any arbitrary level $k$ starting from \emph{null} diagram ($\emptyset$), means \emph{no box}. For a restricted growth process, one can assign a probability to every transition. Denoting a particular Young diagram at level $k$ by $\lambda_k$ we associate a transition probability $\cP_{\text{transition}}(\lambda_k,\lambda_{k+1})$ for a transition from $\lambda_k$ to $\lambda_{k+1}$
\ben\label{eq:transmeasure}
\cP_{\text{transition}}(\lambda_k,\lambda_{k+1}) = \frac{1}{k+1} \frac{\text{dim}\  \lambda_{k+1}}{\text{dim} \ \lambda_{k}}
\een
if $\lambda_{k+1}$ is obtained from $\lambda_k$ by adding one box, otherwise $\cP_{\text{transition}}(\lambda_k,\lambda_{k+1})=0$. A growth process, following the above probability measure, is called \emph{Plancherel growth process} (see \cite{kerov_book,Hora:2237389} for a comprehensive review). Note that the probability to get a diagram at level $k+1$ from a diagram at level $k$ does not depend on the history of transition from level $k-1$ to $k$. Thus, the growth process is Markovian. It was shown by Vershik and Kerov \cite{VerKer77} and independently by Logan and Shepp \cite{LogShe} that Young diagrams following Plancherel growth process converge to a \emph{universal diagram} in the large $k$ limit when the diagrams are normalised (scaled) appropriately such that the area of the diagram is unity. The boundary of such normalised diagram becomes smooth under scaling. A universal Young diagram means the boundary curve takes a particular form, which is called the \emph{limit shape}. The limit shape follows the famous \emph{arcsin law} \cite{Kerov1,VerKer77}. In the continuum (large $k$) limit Kerov introduced a differential model to capture the growth of Young diagrams \cite{Kerov1,kerov_book}. He associated a \emph{`time'} parameter with the continuous diagrams to study the evolution of these diagrams. It turns out that Young diagrams equipped with Plancherel measure follow an evolution equation which is a first order partial differential equation. The model was named as \emph{automodel} \cite{Kerov1,kerov_book}. The class of diagrams following such a growth or evolution equation is called \emph{automodel class}. The limit shape is a unique solution of the automodel equation in far future with $\emptyset$ as initial condition in far past.

Random matrix models offer an independent way to analyse the growth processes of Young diagrams. Partition function of unitary matrix models can be written as a sum over representations of the unitary group \cite{Kazakov:1995ae, duttagopakumar}. In the large $N$ limit, one can rescale the Young diagrams appropriately and find the representations that dominate the partition function via saddle point analysis \cite{duttagopakumar}. These dominant representations are indeed asymptotic Young diagrams. Growth of such continuous representations depends on the action and parameters of the model under consideration. {However, such problems demands further investigation and study which to the best of our knowledge is lacking either in mathematics or physics literature}\footnote{A matrix model analysis of such growth process has been discussed in \cite{Eynard:2008mt} - an asymptotic shape of large diagram has been computed which matches with the limit shape of \cite{2003math.4010I,Kerov1,LogShe,VerKer77}.}. \emph{In this paper we try to understand whether one can write down a unitary matrix model which can describe the asymptotic growth of Young diagrams.} To our surprise, we find that the Plancherel growth process is actually captured by the simple \emph{Gross-Witten-Wadia} (GWW) model. GWW model is one of the well studied and exactly solvable model in unitary random matrix theory. It is therefore quite interesting to see that the strong coupling (no-gap) phase of GWW model is also {capable of capturing} the Plancherel growth of Young diagrams. In fact we show that the matrix model computation offers a direct and simple proof of the limit shape theorem of Kerov. In this paper, we go further and map the growth process with the evolution of free Fermi droplets in 2 dimensions in the context of unitary matrix model \cite{duttagopakumar}. By quantising these droplets we also construct a Hilbert space associated with such a growth process and show that different diagrams in the automodel class are mapped to \emph{coherent} states in the Hilbert space. Therefore, our analysis maps the growth process of Young diagrams to the evolution of \emph{classical} or coherent states in the Hilbert space in the large $N$ limit. Large $N$ fluctuations about the automodel diagrams \cite{2003math.4010I} are mapped to quantum fluctuations of coherent states.

The salient observations of this paper are following.
\begin{itemize}
    \item Following \cite{Eynard:2008mt} we compare the growth of Young diagrams with that of 2-dimensional crystals. We consider ensemble of Young diagrams with infinite members. Assigning Plancherel probability for a particular diagram we write down a grand canonical partition function for such an ensemble.
    \item The partition function can be computed exactly if we sum over all possible representations of symmetric group $S_k$ for a given box number $k$. However that does not seem to be very exciting. To capture the growth process from the partition function we impose a \emph{cut-off} $N$ on the number of rows of the Young diagrams. This essentially restricts us to a sub-ensemble where all the diagrams have maximum $N$ number of rows. As a benefit, we are now allowed to study the system under saddle point approximation in the large $N$ limit.
    \item We next show that the above partition function is same as the partition function of $SU(N)$ GWW model where the rank of the gauge group is same as the cut-off. We use the Frobenius formula to establish the equality.
    \item We observe that the no-gap phase of GWW model is mapped to the automodel class of Kerov. The coupling constant of GWW model plays the role of \emph{time} in automodel class. The limit shape corresponds to the GWW transition point. 
    \item We next focus on the free Fermi droplet description of the growth process. Large $N$ phases of Unitary matrix models admit a droplet description \cite{duttagopakumar}. We quantise these droplets and construct the Hilbert space. We show that different Young diagrams in automodel class are mapped to coherent states in the Hilbert space. Evolution of Young diagrams are, therefore, mapped to evolution of these coherent states. Thus our analysis explicitly maps the Plancherel growth to the evolution of classical states in the Hilbert space.
    \item The Hilbert space description of the growth process also allows us to map the large $N$ fluctuations about the automodel diagrams to the quantum fluctuations of coherent states. In this paper we explicitly discuss this mapping.
    
\end{itemize}

The plan of the paper is following. In Section \ref{sec:umm}, we review earlier works where we map a particular class of unitary matrix models to two dimensional droplets made of fermions in phase space. Subsequently in Section \ref{sec:plancherel}, we discuss Plancherel growth of Young diagrams in details following the works of \cite{VerKer77}. The growth process is mapped to the evolution of a grand canonical ensemble in Section \ref{sec:partitionfunction}. Section \ref{sec:largek} is devoted to the analysis of the partition function that we set up for the Plancherel growth process of Young diagrams. We provide an alternate proof of the \emph{limit shape theorem} of Kerov and Vershik. In Section \ref{sec:hilbertspace}, following a quantization procedure, we provide a Hilbert space description of the growth process. \ref{app:ummlysis} provides some details of the analysis of a generic class of UMMs in terms of eigenvalues analysis as well as Young diagrams. It further discusses a connection between the two pictures. In \ref{app:speccurve} we discuss the spectral curve for the no-gap solution of the unitary matrix model. Finally, \ref{app:asymetricsol} discusses an asymmetric solution that comes up in the context of GWW model which has connections with the growth process that we discuss in this work.


\section{Review 1 : Droplet description of unitary matrix models }
\label{sec:umm}

We start with a very brief discussion on droplet description of different large $N$ phases of a generic UMM. The partition function of UMM is given by,
\begin{equation}\label{eq:Zumm}
Z=\int [dU] e^{S(U)}
\end{equation}
where $[dU]$ is the Haar measure and $S(U)$ is a generic function of $N\times N$ unitary matrices $U$. The Haar measure $[dU]$ and $S(U)$ are invariant under unitary transformation. A particular class, but yet general, of unitary matrix models is given by the following action 
\ben\label{eq:partfuncplaq}
S(U)=N \sum_{n=1}^{Q}\frac{\beta_n}{n}\lb \Tr U^n +\Tr U^{\dagger n}\rb 
\een
where $Q$ is any arbitrary positive integer and $\beta_n$s are some arbitrary real parameters. This model is called the \emph{single plaquette model}. This model has applications in lattice gauge theory. Partition functions for different large $N$ Chern-Simons matter theories can also be written as a single plaquette model \cite{Chattopadhyay:2019lpr}. The large $N$ phase structure of (\ref{eq:partfuncplaq}) for $Q=2$ has been studied in \cite{zalewski,Mandal:1989ry}. A special case of single plaquette model is Gross-Witten-Wadia model \cite{gross-witten,wadia}, where $\beta_{n>1}=0$. {The GWW model is one of the exactly solvable models and it appears in many contexts in physics as well as in mathematics.}

Since both the Haar measure and the action in equation (\ref{eq:Zumm}) are invariant under unitary transformations, one can go to a diagonal basis
$$
U\ra \{e^{i \theta_1}, \cdots e^{i \theta_i}, \cdots, e^{i\theta_N}\},
$$
where $\theta_i$s are the eigenvalues of $U$. In the large $N$ limit a particular distribution of these eigenvalues on unit circle dominates the partition function. Defining a distribution function (called eigenvalue density) $\rho(\theta)$
\ben\label{eq:evdensity}
\rho(\theta) = \frac1N \sum_{i=1}^N\delta(\theta-\theta_i) = \frac{dx}{d\theta}
\een
one can show that in the large $N$ limit the eigenvalue density satisfies the saddle point equation
\ben\label{eq:evsaddle2}
\Xint- d\theta' \cot\lb \frac{\theta-\theta'}{2}\rb \rho(\theta') = \sum_n \beta_n \cos n\theta, \quad \with \quad \int d\theta \rho(\theta) = 1.
\een
This equation is obtained by extremising the effective action associated with the partition function (\ref{eq:Zumm}). For a given set of parameters $\{\beta_n\}$ one can solve this equation to find $\rho(\theta)$ corresponding to different large $N$ phases of the model under consideration. Different large $N$ phases are classified by number and positions of gaps in $\rho(\theta)$ \cite{BIPZ,zalewski,Mandal:1989ry,gross-witten,wadia} on a unit circle.

A unitary matrix model can be equivalently analysed in terms of $SU(N)$ (or $U(N)$) representations \cite{douglas2,duttagopakumar,duttadutta,Chattopadhyay}, which are expressed in terms of Young diagrams and hence the generic partition function (\ref{eq:Zumm}) can be written in Young diagram basis as well. Expanding the partition function in representation basis one finds
\ben\label{eq:pffinal2} 
\cZ=\sum_{\{h_i\}} \sum_{\vec k, \vec l}
\frac{\varepsilon(\vec \b, \vec k) \varepsilon(\vec \b, \vec
	l)}{z_{\vec k} z_{\vec l}} \chi_{\vec h}(C(\vec k)) \chi_{\vec
	h}(C(\vec l)) \quad \where, \quad \varepsilon(\vec \b, \vec k) =
\prod_{n=1}^{\infty}(N\b_{n})^{k_n}, \ \
z_{\vec k} = \prod_{n=1}^{\infty} k_{n}! n^{k_n}.
\een
Here, $\{h_1,\cdots,h_N\}$ are a set of hook numbers characterising a Young diagram, $\chi_{\vec h}(C(\vec k))$ is the character of the permutation group $S_k$ for a conjugacy class $C(\vec k)$ with $k=\sum_n n k_n$ in representation $\{h_i\}$. In the large $N$ limit the partition function (\ref{eq:pffinal2}) is dominated by a Young diagram which is characterised by a density function $u(h)$
\be\label{eq:uhdef}
u(h) = - {d x\over d h}.
\ee
Since $h(x)$ is a monotonically decreasing function of $x$, Young density has an upper and lower cap $1\geq u(h) \ge 0 \ \forall \ x\in[0,1]$. Large $N$ phases are characterised by different $u(h)$. It is difficult to find a saddle point equation for $u(h)$ for a generic plaquette model (\ref{eq:partfuncplaq}), but one can reconstruct dominant Young diagrams \cite{Chattopadhyay} through the droplet picture. See \ref{app:YDalysis} for details.

\subsection{The large $N$ droplets}
\label{sec:connection}

Droplet description of large $N$ phases is based on the fact that the partition function of UMM has two equivalent descriptions - in eigenvalue basis and in Young diagram basis. It is well known that eigenvalues of unitary matrices behave like position of free fermions \cite{BIPZ}. On the other hand, hook lengths of Young diagrams are like momenta of these fermions \cite{duttagopakumar,douglas2}. A relation between these two pictures offers droplet or phase space description for different classical phases \cite{duttagopakumar,duttadutta,Chattopadhyay}.

It was first observed in \cite{duttagopakumar} that the eigenvalue and the Young diagram distributions are functional inverses of each other for a particular class of unitary matrix model, namely the $(a,b)$ model whose phase structure is similar to that of GWW model. Later in a series of papers \cite{duttadutta,riemannzero,Chattopadhyay} such relations were proved for a generic class of UMM. The details of the calculation is given in \ref{app:connection}. Here we present the main result.

In the large $N$ limit, the connection between eigenvalue density $\rho(\theta)$ and hook number $h$ is given by
\ben\label{eq:hrhorel}
h^2-2S(\theta) h + S^2(\theta)-\pi^2 \rho^2(\theta) =0
\een
where 
\ben\label{eq:Stheta}
S(\theta)=\frac12+\sum_n \beta_n \cos n\theta.
\een
Using this equation one can also define a spectral curve for the generic UMM. See \ref{app:speccurve}.

The relation (\ref{eq:hrhorel}) allows us to provide a phase space picture for different large $N$ phases of UMM in terms of free fermi droplets (two dimensional distributions). These distributions/droplets are similar to Thomas-Fermi distributions \cite{thomasfermi1,thomasfermi2}. Equation (\ref{eq:hrhorel}) has two possible solutions $h_{\pm}(\theta)$ given by
\ben\label{eq:hpm1}
h_{\pm}(\theta)=S(\theta)\pm \pi \rho(\theta).
\een
Using this relation we define a distribution function $\omega(h,\theta)$ in $(h,\theta)$ plane
\be \label{eq:phasespacedistri}
\omega(h,\q) = \Theta\lb{(h-h_-(\q))(h_+(\q)-h)\over 2}\rb
\ee
such that $\o(h,\q)=1$ for $h_-(\q)<h<h_+(\q)$ and zero otherwise.
Eigenvalue distribution, by construction, can be obtained by integrating
out $h$ for a given $\q$ and is given by
\be\label{eq:rhofromphasespace}
\r(\q)= \frac1{2\pi}\int dh \ \omega(h,\q) = {h_+(\q)-h_-(\q)\over 2\pi}.
\ee
The distribution also satisfies the normalisation condition
\ben\label{eq:omeganormalisation}
\frac1{2\pi}\int dh \ d\theta \ \omega(h,\theta) =1.
\een
The function $S(\q)$ can also be written in terms of phase space geometry
\be\label{eq:Sfromphasespace}
S(\q) = \frac1{2\pi \r}\int_0^\infty dh \ h \ \o(h,\q) =
{h_+(\q)+h_-(\q) \over 2}.
\ee
Integrating $\omega(h,\theta)$ over $\theta$ for a given $h$ gives a distribution of hook numbers $h$
\ben\label{eq:hdistribution}
w(h) = \frac1{2\pi}\int d\theta \ \omega(h,\theta)\quad \with \quad \int dh \ w(h) =1.
\een
$w(h)$ captures the information about large $N$ representations. Since $\omega^2(h,\theta) = \omega(h,\theta)$, it is actually the shape (i.e. boundary) of this distribution function which captures information about different large N phases of the theory. We call such two dimensional distributions large $N$ droplets.

The matrix model, we are considering, has no dynamics. Different large $N$ phases, obtained by solving the saddle point equation, are possible minimum free energy configurations of the model. Hence the corresponding droplets are static - the shapes are not changing with time. To incorporate dynamics into the picture, we follow an \emph{ad hoc} way : identify large $N$ droplets with Thomas-Fermi distribution at zero temperature and obtain the single particle Hamiltonian by comparing the two\footnote{One can also construct similar droplet description for time dependent unitary matrix model. In that case time evolution of these droplets are inherent \cite{Chattopadhyay:2020rle}}. In this way it is possible to incorporate time in our system.

Comparing the distribution function (\ref{eq:phasespacedistri}) with Fermi distribution at zero temperature
\be
\Delta(p,q) = \Theta(\mu - \mathfrak{h}(p,q))
\ee
($\mu$ is chemical potential and $\mathfrak{h}(p,q)$ is single particle Hamiltonian density) we find that the single particle Hamiltonian density is given by
\ben\label{eq:singleh}
\mathfrak{h}(h,\q)= {h^2\over 2} - S(\q)h + {g(\q)\over 2} + \mu, \quad 
\where \quad g(\q) = h_+(\q) h_-(\q).
\een
Total Hamiltonian\footnote{One can show that \cite{Chattopadhyay} integrating over $h$, the total Hamiltonian (without the $\hbar$ factor) is same as the collective field theory Hamiltonian of Jevicki and Sakita \cite{jevicki} 
\ben
\begin{split}\label{eq:Hamcoll}
H_h & = - \int d\q \lb \frac{S^2 \r}2 + \frac{\pi^2 \r^3}6 + V_{eff}(\theta)\rho\rb 
+\m
\end{split}
\een
with an effective potential.} can be obtained by integrating $\mathfrak{h}(h,\q)$ over the phase space
\ben\label{eq:Hh}
\begin{split}
	H_h &=\frac1{2\pi \hbar} \int d\q 	\int dh \ \o(h,\q) \ \mathfrak{h}(h,\q).
\end{split}
\een
We have taken into account the fact that one state occupies a phase space area of $2\pi \hbar$ in semi-classical approximation. It needs to modify the normalisation of phase space density
\ben
\frac{1}{2\pi \hbar}\int dh d\theta \omega(h,\theta) = N, \quad \with \quad  \hbar N =1 
\een
where, $N$ is total number of states available inside a droplet. The classical limit corresponds to $\hbar\ra0, \ N\ra \infty \ \with \  \hbar N =1$.


\section{Review 2 : Plancherel Growth of Young diagrams}
\label{sec:plancherel}

The Markovian growth process of large Young diagrams can be given a statistical interpretation described through a partition function. In order to achieve that we define a Young lattice \cite{Eynard:2008mt}
\be\label{eq:Ylattice}
\cY = \bigcup\limits_{k=0}^{\infty}  \cY_k.
\ee
All the members of $\cY_k$ have same number of boxes but different shapes. $\cY_k$ can be thought of as an ensemble of Young diagrams with the same macroscopic variable $k$. Hence, we can define a partition function for the grand canonical ensemble (\ref{eq:Ylattice})
\ben\label{eq:GCPF}
\cQ_\cY = \sum_{k=0}^{\infty}z^k  \cZ_{\cY_k} 
\een
where $z$ is called \emph{fugacity} ($z>0$) and $\cZ_{\cY_k}$ is the canonical partition function for $\cY_k$, given by
\be\label{eq:CPF}
 \cZ_{\cY_k} = \sum_{\lambda_k} \cP(\lambda_k)  \delta(k-|\lambda_k|).
\ee
$\cP(\lambda_k)$ is the probability associated with the ensemble member $\lambda_k$ - a Young diagram with $k$ number of boxes.

To make the growth process meaningful it is customary to assign a probability for each diagram at level $k$ in the Young lattice $\cY$. There is a natural way to assign probability to different diagrams. We count the total number of inequivalent paths one can follow to come to a particular diagram at level $k$ starting from $\emptyset$. See figure \ref{fig:younggrowth}. It turns out that the Plancherel measure is proportional to the square of that number. The proportionality constant is fixed by the normalization condition. To calculate the number of paths heading to a diagram $\lambda_k$ we look at growth of Young tableaux rather than Young diagrams. Starting from $\emptyset$ we keep on adding one box at each level with increasing number. Therefore the readers can easily convince themselves that at each level $k$ we have different Young tableaux and a particular tableaux can be reached from $\emptyset$ by a unique path only. Thus the number of paths available to reach a particular Young diagram $\lambda_k$ is equal to the number of standard Young tableaux $f_{\lambda_k}$ of that given shape. It is well known that \cite{hamermesh,fulton1991representation},
\be\label{eq:dimRnormalization}
\sum_{\lambda_k \in\cY_k} (f_{\lambda_k})^2=k!.
\ee
Hence we get the Plancherel measure $\cP(\lambda_k)$ for a diagram $\lambda_k$
\be\label{eq:Plancherelmeasure}
\cP(\lambda_k) =  \frac{f_{\lambda_k}^2}{k!}.
\ee
This definition of Plancherel probability is equivalent to what we defined in equation (\ref{eq:transmeasure}). One can show that
\ben
\cP(\lambda_k)=\sum_{\text{paths}}\prod_{i=0}^{k-1} \cP_{\text{transition}}(\lambda_i,\lambda_{i+1}).
\een

We use $\cP(\lambda_k)$ to write the partition function for the growth process. The number $f_{\lambda_k}$ is equal to the dimension of the representation $\lambda_k$, i.e.
\be
f_{\lambda_k} = \text{dim} \ \lambda_k.
\ee
and thus we have,
\be\label{eq:Plancherelmeasure2}
\cP(\lambda_k) =  \frac{(\text{dim} \ \lambda_k)^2}{k!}.
\ee
It was observed in \cite{LogShe,VerKer77} that a Young lattice equipped with Plancherel measure terminates to a universal class of diagram in the limit $k\ra \infty$ when the diagrams are scaled properly.
 
\subsection{The Universal Diagram and Automodel}
\label{sec:automodel}

\begin{figure}[h]
	\begin{center}
		\includegraphics[scale=0.2]{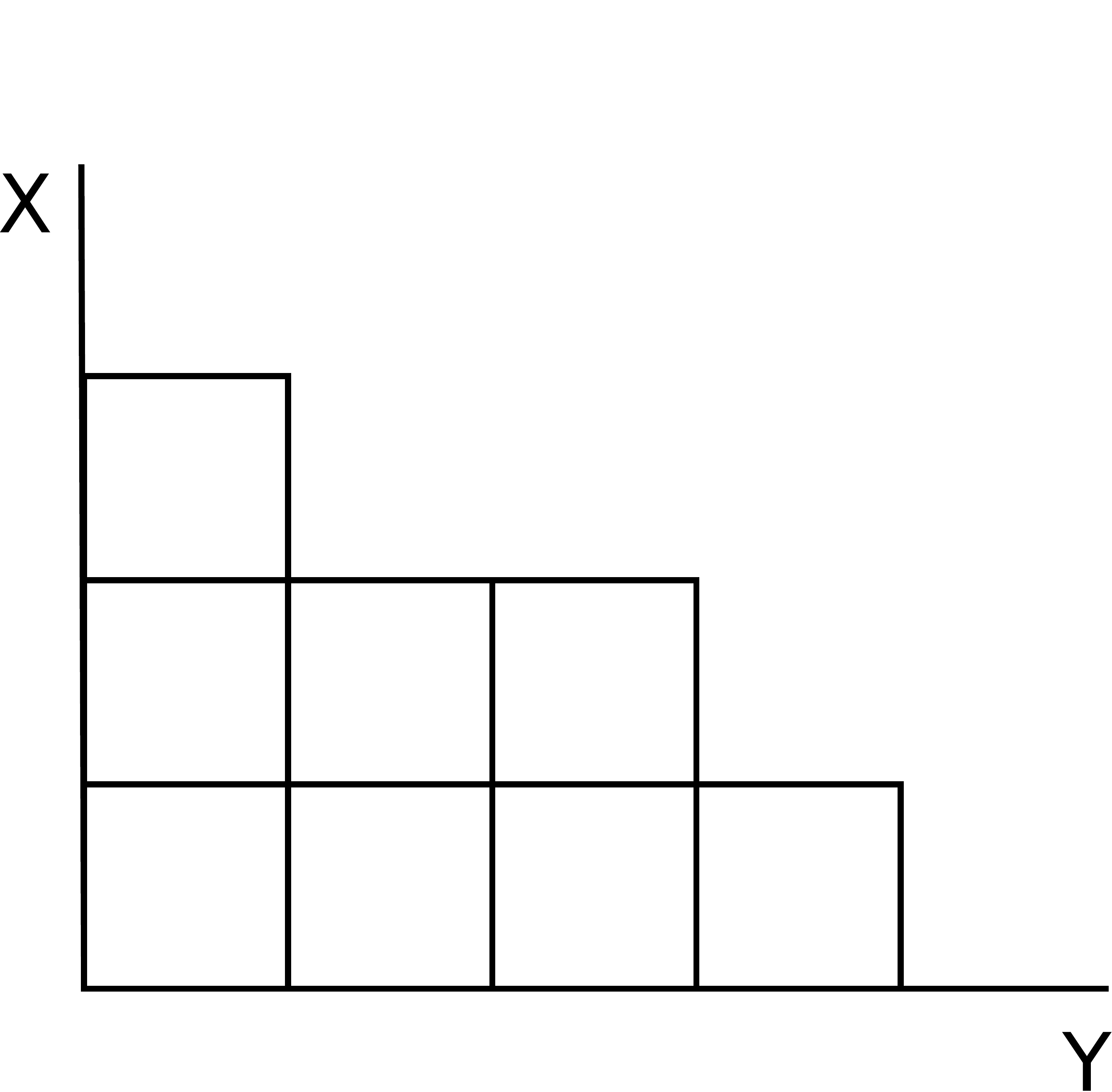}
		\caption{\footnotesize{Typical structure of a Young diagram in French notation.}}
		\label{fig:YDFrench}
	\end{center}
\end{figure}

\begin{figure}[h]
	\begin{center}
		\includegraphics[scale=0.17]{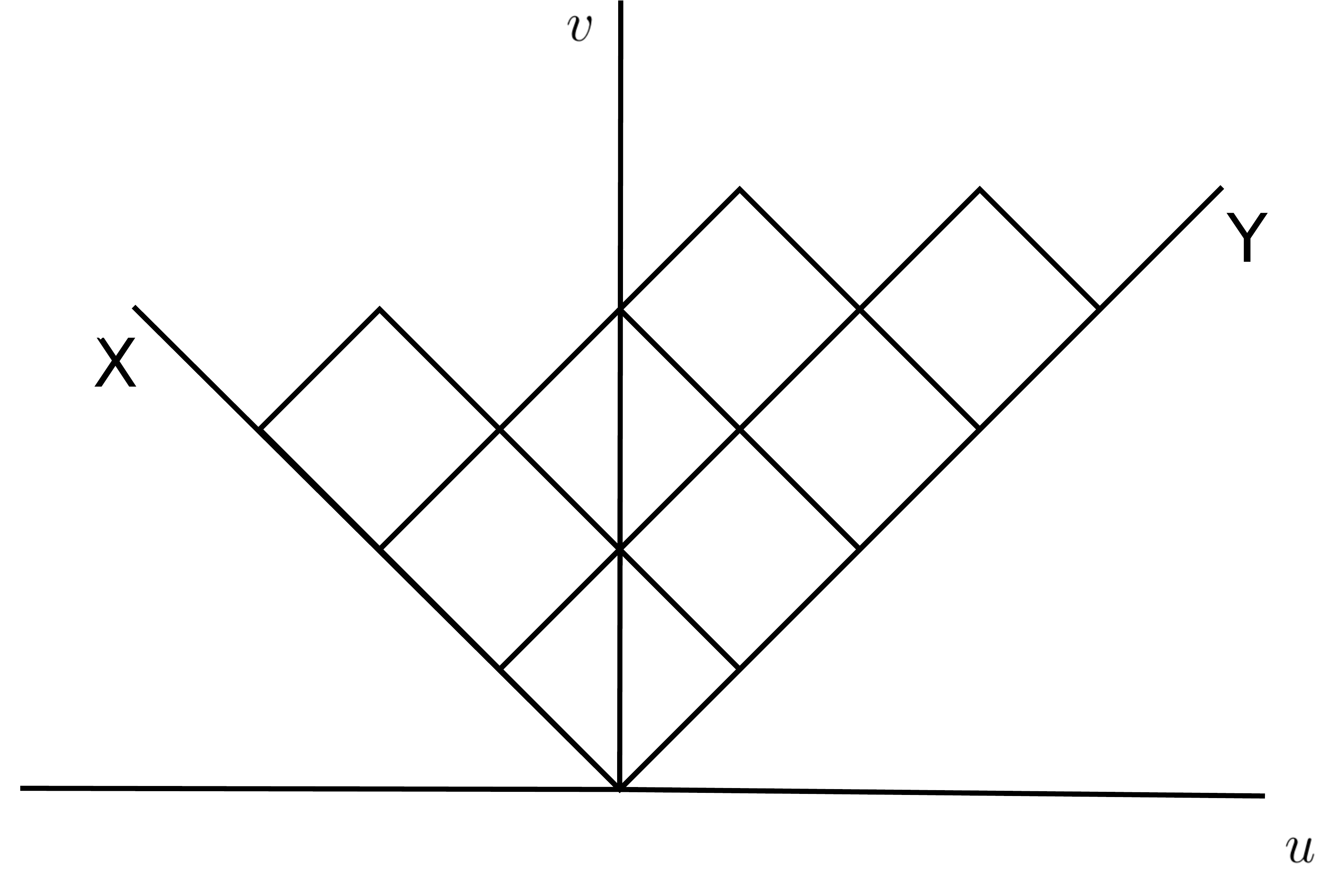}
		\caption{\footnotesize{Typical structure of a $45^{\circ}$ anti-clockwise rotated Young diagram.}}
		\label{fig:YDFrench-rotated}
	\end{center}
\end{figure}

Although we are using the ``English" notation for Young diagrams, but the limit shape of Young diagrams takes a simple form in rotated French notation. A typical shape of Young diagram in French notation is given in \cref{fig:YDFrench}. The centres of boxes are marked with $(X,Y)$ coordinates. The function $X(Y)$ specifies a particular shape of Young diagram in this notation.
However, it is more convenient to rotate this diagram anti-clock wise by $\pi/4$ and work in the redefined coordinates
\begin{equation}
u=\frac{1}{2}(Y-X) \qquad v=\frac{1}{2}(Y+X).
\end{equation}
A Young diagram in this notation\footnote{There is another advantage to draw the Young diagrams in rotated French notation. A transition from $\lambda_k$ to $\lambda_{k+1}$ occurs when one keeps a box at any of the minima of rotated diagram. Putting a box at different minima corresponds to different diagrams at $k+1$ level. Therefore the transition probabilities are denoted by $\mu_a$ where $a$ is the position of a minimum. To find $\mu_a$ we define two polynomials $P(x) = \prod_{a=1}^{n}(x-x_a)$ and $Q(x)= \prod_{a=1}^{n-1}(x-y_a)$, where $x_1,\cdots , x_n$ are positions of consecutive minima and $y_1,\cdots,y_{n-1}$ are consecutive maxima. The transition probability from $\lambda_k$ to $\lambda_{k+1}$ by adding a box at $a^{th}$ minima is given by decomposing the quotient into partial fraction
\ben
\sum_{a=1}^{n}\frac{\mu_a}{x-x_a} = \frac{Q(x)}{P(x)}.
\een
}
is depicted in figure \ref{fig:YDFrench-rotated}.
Note that here the coordinate $u$ is different than the Young diagram density defined in (\ref{eq:uhdef}). For a finite number of boxes the function $v(u)$ is rough and zig-zag i.e. $v'(u)=\pm 1$. As the number of boxes goes very large we define a rescaled function 
\be\label{eq:vk}
\hat v_k(u) = \frac{v(u\sqrt{k})}{\sqrt{k}}
\ee
such that the area under the curve is finite and the boundary curve becomes smooth. It was observed by \cite{LogShe,VerKer77} that when the growth process follows Plancherel transition probability (\ref{eq:transmeasure}) the asymptotic shape of rescaled Young diagrams converges uniformly to a unique curve given by
\begin{equation}
\label{eq:kerovfunction}
\lim_{k \ra \infty}\hat v_k(u)\equiv\Omega(u)=
\begin{cases}
\frac{2}{\pi}(u \sin^{-1} \frac{u}{2}+\sqrt{4-u^2}) &\mbox{if} \quad |u| \leq 2\\
|u|& \mbox{if} \quad |u|>2.
\end{cases}
\end{equation}

In the continuum limit Kerov introduced \cite{Kerov1} \emph{charge} of a diagram, denoted by $\sigma(u)$ and is given by (we are using the notation that $\hat v(u) =\hat v_k(u)$ in the large $k$ limit)
\ben\label{eq:youngcharge}
\sigma(u) = \frac12 \lb\hat v(u)-|u|\rb.
\een
Therefore,
\ben\label{eq:chargeprime}
\sigma'(u) =\bigg\{ {\ \ \frac12 +\frac{\hat v'(u)}{2} \quad \for \quad u<0 \atop -\frac12 +\frac{\hat v'(u)}{2} \quad \for \quad u>0}.
\een
One can define moments of a diagram
\ben\label{eq:moments}
p_n = -n \int u^{n-1} d\sigma(u)
\een
such that area of a diagram (area covered under the curve $\hat v(h)$) is given by $A=(p_2-p_1^2)/2$. It is convenient to consider a moment generating function
\ben\label{eq:momgenfunc}
S(x) = \sum_{n=1}^{\infty}\frac{p_n}{n}x^{-n} = \int \frac{d\sigma(u)}{u-x}.
\een
The moment generating function as well as the sequence of moments determine the charge and hence the diagram ($\hat v(u)$) completely. The moment generating function plays an important role in our large $k$ analysis of partition function.

In \cite{Kerov1} Kerov introduced a dynamical model for the growth of Young diagrams. For every continuous Young diagram characterized by the function $\hat v(u)$, one can define the function $v(u,t)$, called the automodel tableaux which depends on two variables $u$ and $t$ as
\begin{equation}
\hat v(u,t)=\sqrt{t} \ \hat v(u/\sqrt{t}) \quad \mbox{for} \quad t>0\ .
\end{equation}
Kerov showed that the Young diagrams, following Plancherel growth, belong to \emph{automodel} class and satisfy the equation
\begin{equation}\label{eq:automodeleqnv}
{\partial_t \hat v(u,t)}=\frac{1}{2t}(\hat v(u,t)-u\partial_u \hat v(u,t)).
\end{equation}
In terms of charges the automodel equation is given by,
\ben\label{eq:automodeleqnsigma}
\dow_t \sigma'(u,t)+\frac{u}{2t} \sigma^{''}(u,t)=0.
\een

\subsection{Fluctuations of automodel diagrams}

Fluctuations of Young diagrams have been under investigation in the mathematics literature \cite{Kerovfluctuation,johansson1998,2003math.4010I,2005math.1112S,2006math.7635B,BogachevSu,2014arXiv1402.4615D,2016arXiv160805163E,2017arXiv170402352D,2004math.5191M,2004math.5258M,2006math.6431C}. A \emph{rescaled} Young diagram defined in (\ref{eq:vk}) in the $k\ra \infty$ limit takes the form of limit shape (denoted by $\Omega$, as defined in \eqref{eq:kerovfunction}). However there can be large $k$ corrections to this result and we call such corrections as fluctuations of limit shape diagram $\Omega$. Kerov studied the Gaussian fluctuations around the limit shape of Young diagrams $\Omega$ endowed with Plancherel measure in \cite{Kerovfluctuation}. In \cite{2003math.4010I}, Ivanov and Olshanki reconstructed a proof of Kerov's result on fluctuations around the limit shape from his unpublished work notes, 1999. The central result pertains to large $k$ corrections to the limit shape which can be stated as
\begin{equation}\label{eq:flucLS}
\lim_{k \to \infty} \hat{\nu}_k(u) \sim \Omega(u)+\frac{2}{\sqrt{k}}\Delta(u)
\end{equation}
The sub-leading piece $\Delta(u)$ is a \emph{Gaussian process} defined for $|u|\leq 2$. More precisely, $\Delta(u)$ is a random trigonometric series given by
\begin{equation}\label{eq:deltafluc}
\Delta(u)=\Delta(2\cos \theta)=\frac{1}{\pi}\sum_{n=2}^{\infty}\frac {\alpha_n}{\sqrt{n}}\sin(n\theta)\ ; \qquad u=2\cos \theta
\end{equation}
where $\alpha_n$ are independent Gaussian random variables with mean 0 and variance 1. Further investigations has been done towards understanding the central limit theorem for Gaussian fluctuations around the limit shape \cite{2006math.7635B,BogachevSu}. Fluctuations of random Gaussian and Wishart matrices have been related to the notion of free probability and free cumulants in earlier works \cite{2004math.5191M,2004math.5258M,2006math.6431C}.

\section{Partition Function for Young lattice}\label{sec:partitionfunction}

Following (\ref{eq:GCPF}, \ref{eq:CPF}) and \cref{eq:Plancherelmeasure}, the grand canonical partition function for $\cY$ can be written as,
\ben\label{eq:GCPF2}
\cQ_\cY = \sum_{k=0}^{\infty} z^k \sum_{\lambda_{k}} \frac{(\text{dim}\lambda_{k})^2}{k!} \delta(k-|\lambda_{k}|), \quad z>0.
\een
This partition function is related to \emph{Poissonised Plancherel measure} \cite{1999math.5032B}. The above ensemble sometimes is known as \emph{Meixner} ensemble in literature \cite{1999math.6120J,2010ArM.48.79J}.

The above partition function does not capture the growth process of Young diagrams as it can exactly be calculated using the normalization \cref{eq:dimRnormalization}
\be
\cQ_{\cY} = \frac{1}{1-z}.
\ee
However, to study the growth process through the partition function we regularise the sum by imposing a \emph{cut-off} on the Young diagrams in the summation over $\lambda_k$ in eq. (\ref{eq:GCPF2}). We introduce a large positive integer $N$ and constrain that the Young diagrams in lattice $\cY$ can not have more than $N$ rows. In presence of such a cut-off the summation over $\lambda_k$ for $k>N$ is not equal to unity anymore. The \emph{regularised} partition function is therefore given by
\ben\label{eq:GCPF3}
\cQ_\cY^N = \sum_{k=0}^{\infty} z^k \sum_{n_1,\cdots, n_N\geq 0} \frac{(\text{dim}\lambda_{k})^2}{k!} \delta\lb k-\sum_{i=1}^N n_i\rb, \quad \with \quad n_1\geq n_2\geq \cdots\geq n_N\geq 0. 
\een
This regularisation gives us a handle on the partition function to carry out a saddle point analysis of the problem. In the large $N$ limit, the partition function is dominated by a single Young diagram for a given value of $z$. It turns out that the shape of this dominant diagram falls into the automodel class of Kerov \cite{Kerov1}, paramatrised by $z$. For a particular value of the parameter $z$ it becomes the limit shape \cite{VerKer77,LogShe,Kerov1}. Before we present the calculation, we show that the partition function (\ref{eq:GCPF3}) is equivalent to the partition function of $SU(N)$ Gross-Witten-Wadia model and its cousins.

\subsection{A connection between Young lattice and Gross-Witten-Wadia model}

The Gross-Witten-Wadia model is a well studied unitary matrix model in physics. The partition function for this model is defined over an ensemble of $N\times N$ unitary matrices with a real potential $\Tr U+\Tr U^{\dagger}$, where the $trace$ is taken over fundamental representations. The partition function of GWW model is given by
\be\label{eq:GWWPF}
\cZ_{GWW} = \int [dU] \ e^{ \frac N{\lambda} (\Tr U+\Tr U^{\dagger})}, \quad \lambda\geq 0 .
\ee
Gross and Witten \cite{gross-witten} (and independently, Wadia \cite{wadia}) studied this matrix model in the context of lattice QCD and observed that the system undergoes a third order phase transition at $\lambda=2$. Different phases of this model are characterised by the topology of distribution of eigenvalues of the unitary matrix $U$ on a unit circle. The strong coupling phase ($\lambda>2$) corresponds to a gap-less distribution of eigenvalues whereas weak coupling phase ($\lambda <2$) shows a finite gap in eigenvalue distribution. 

A close cousin of GWW model \cite{Aharony:2003sx,AlvarezGaume:2005fv} is given by
\be\label{eq:cZ}
\cZ_{c} = \int [dU] \ e^{a \Tr U\Tr U^{\dagger}}.
\ee
The phase structure and eigenvalue distributions of this model are similar to those of GWW up to a redefinition of parameters : $a \langle \Tr U\rangle = N/\lambda$ \cite{duttadutta}. Expanding the exponential in (\ref{eq:cZ}) we get
\be
\cZ_{c} = \int [dU] \ \sum_{k=0}^{\infty} \frac{a^k}{k!} (\Tr U)^k (\Tr U^{\dagger})^k.
\ee
Using Frobenius formula for the characters of symmetric group we can write
\be
(\Tr U)^k = \sum_R \chi_R(1^k) \Tr_RU, \quad\text{and} \quad (\Tr U^{\dagger})^k = \sum_R \chi_R(1^k) \Tr_RU^\dagger
\ee
where the sum is over representations of $U(N)$ (or $SU(N)$) and $\chi_R(1^k)$ is the character of conjugacy class $(1^k)$ of symmetric group $S_k$ in representation $R$. Finally using the normalization condition for the characters of unitary group
\be
\int [dU] \Tr_RU\Tr_{R'} U^{\dagger} = \delta_{RR'}
\ee
we arrive at the final expression for $\cZ_{c}$ written in terms of sum over representations of $U(N)$ \cite{duttagopakumar}
\ben\label{eq:cZ2}
\cZ_{c} = \sum_{k=0}^{\infty} a^k \sum_{R} \frac{(\chi_R(1^k))^2}{k!} .
\een
It is well known that the character of the conjugacy class $(1^k)$ of symmetric group $S_k$ in representation $R$ is equal to the dimension of the representation \cite{hamermesh}
\be
\chi_R(1^k) = \text{dim} R.
\ee
Representations of $U(N)$ can be expressed in terms of Young diagrams. Since $\chi_R(1^k)$ is non-zero only when total number of boxes in the Young diagram is $k$ we have
\be\label{eq:PFcGWR}
\cZ_{c} = \sum_{k=0}^{\infty} a^k \sum_{\lambda_k} \frac{(\text{dim} \lambda_k)^2}{k!} \delta(k-|\lambda_{k}|).
\ee

Thus we see that the partition function for (cousin of) GWW model is same as that of a Young lattice with the coupling constant $a$ playing the role of fugacity. The rank of the gauge group $SU(N)$ in GWW model plays the role of the cut-off $N$ in (\ref{eq:GCPF3}).

\section{Large $N$ analysis of partition function : an alternate proof of the limit shape theorem}\label{sec:largek}

The large $N$ analysis of partition function (\ref{eq:GCPF2}) was explicitly done in \cite{duttagopakumar}. We briefly review the procedure for the readers not familiar with matrix model techniques (for a more comprehensive treatment of matrix models, see \cite{Marino:2005sj,Eynard:2015aea,eynard_book}). To analyse the partition function (\ref{eq:GCPF3}) we  denote a valid Young diagram of symmetric group $S_k$ by a set of numbers $\{n_i\}_{i=1}^N$ where $n_i$ denotes the number of boxes in $i^{th}$ row. 
\begin{figure}[h]
	\begin{center}
		\includegraphics[width=9cm,height=5.5cm]{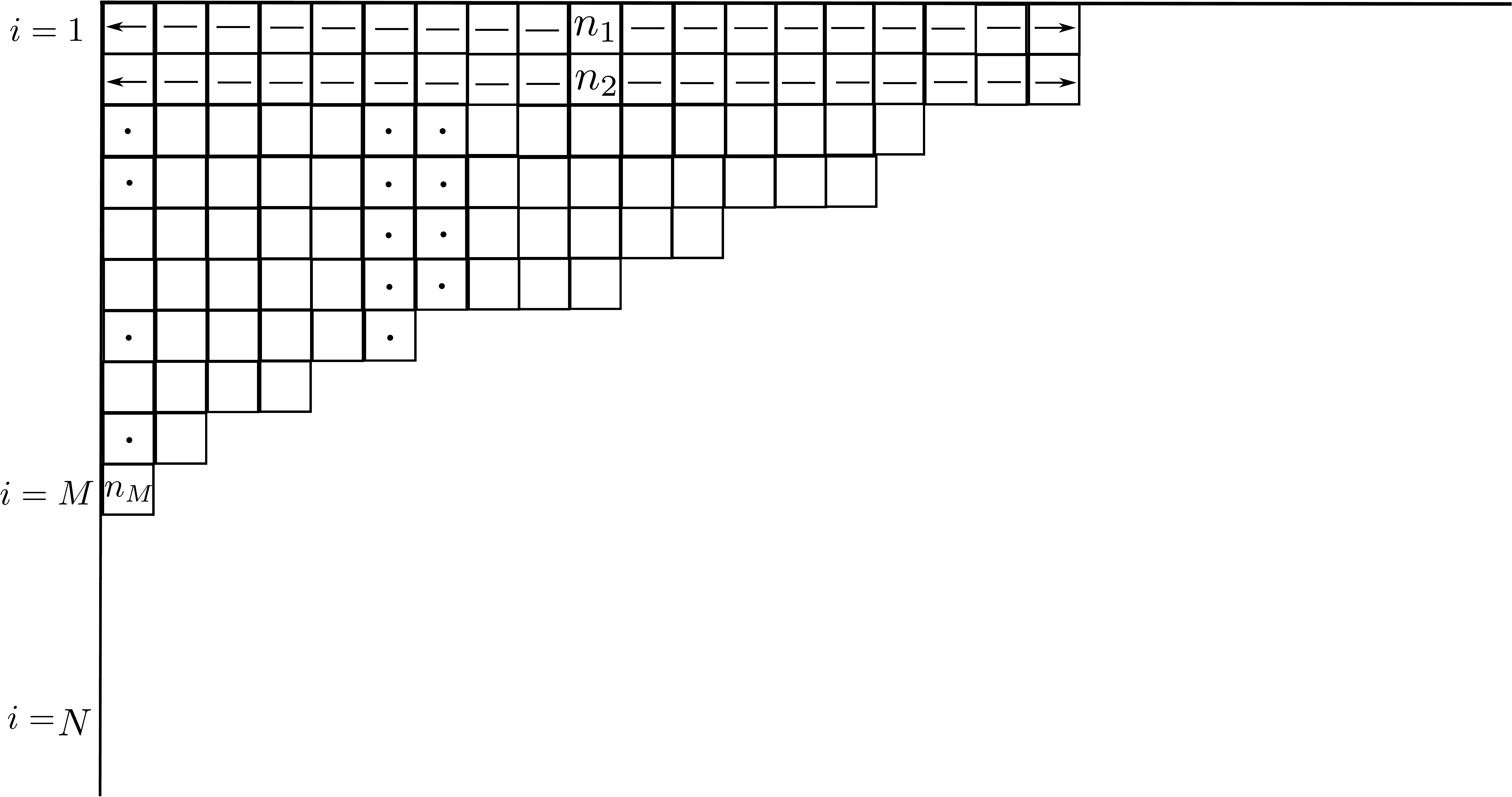}
			\caption{A generic Young diagram in English notation. Here $N$ is an arbitrary positive integer. The number of boxes in the first column is less than or equal to $N$. In general, $\exists$ a number $0<M\leq N$ such that $n_i=0$ for $i=M+1, \cdots, N.$}
			\label{fig:YD-english}
	\end{center}
\end{figure}
$N$ is an arbitrary positive integer greater than or equal to the height of the first column. See figure \ref{fig:YD-english}. The dimension of a representation $\lambda_k$ of $S_k$ is given by \cite{hamermesh}
\ben\label{eq:dimY}
\text{dim} \lambda_k = \frac{k!}{h_1! h_2! \cdots h_N!} \prod_{i=1\atop i<j }^{N}(h_i-h_j)
\een
where,
\ben\label{eq:hook}
h_i = n_i + N -i
\een
is the hook length of the first box in $i^{th}$ row. 

We consider the large $N$ limit of the partition function (\ref{eq:GCPF3}). In this limit the hook numbers $h_i \sim N$ (\ref{eq:hook}). Therefore we define the following continuous functions to describe Young diagrams at large $N$
\ben
n(x) = \frac{n_{i}}{N}, \quad h(x) = \frac{h_i}{N}, \quad \where \quad x=\frac{i}{N},\quad x\in[0,1].
\een
The function $n(x)$ or $h(x)$ captures the distribution of boxes in a large $k$ Young diagram. The relation between $n(x)$ and $h(x)$ follows from equation (\ref{eq:hook}) and is given by
\be\label{eq:hook2}
h(x) = n(x)+1-x.
\ee
The number of boxes in a Young diagram in the large $N$ limit is given by
\be
k = \sum_{i=1}^{N} n_i \longrightarrow N^2 \lB \int_0^1 dx \lb h(x)+1-x \rb\rB = N^2 \lB \int_0^1 dx h(x)-\frac12 \rB = N^2 k'
\ee
where
\be
k' =  \int_0^1 dx h(x)-\frac12\
\ee
is the renormalised box number and is an ${\cal O}(1)$ quantity. Thus we see that the number of boxes in a Young diagram in the large $N$ limit goes as $\sim {\cal O}(N^2)$ and hence $N\sim {\cal O}(\sqrt{k})$. The partition function (\ref{eq:GCPF3}) in $N \ra \infty$ limit is given by,
\ben
	\cQ_{\cY} = \int [Dh(x)] e^{-N^2 \seff{h(x)}} 
\een
where
\ben 
-\seff{h(x)}  = \int_0^1 dx\Xint-_0^1 dy \ln|h(x)-h(y)| -2 \int_0^1 dx h(x) \ln h(x) +k' \ln (z k') +k'+1.
\een
Therefore, the dominant contribution to the partition function comes from the extrema of $\seff{h(x)}$. Varying $\seff{h(x)}$ with respect to $h(x)$ we get the saddle point equation
\ben\label{eq:saddeq2}
\Xint- \frac{u(h')dh'}{h-h'} = \ln \lb \frac{h}{\xi}\rb, \quad \where \quad \xi^2 = z k'
\een
where the Young diagram density $u(h)$ is given by
\ben\label{eq:uhdef}
u(h)=-\frac{\partial x}{\partial h}.
\een
Monotonicity of $h(x)$ implies $0 \leq u(h) \leq 1$. $u(h)$ also satisfies two conditions
\ben\label{eq:uhcond}
\int dh\, u(h) =1, \quad \tand \quad \int h\, u(h)\, dh = k'+\frac12.
\een
We solve the saddle point equation (\ref{eq:saddeq2}) to find Young diagram density that satisfies the constraints (\ref{eq:uhcond}).

\subsection{Large \emph{N} solutions and automodel diagrams} \label{sec:largeNphases}

All possible large $N$ solutions of (\ref{eq:saddeq2}) were discussed in \cite{duttagopakumar} and it was observed that (\ref{eq:saddeq2}) admits two possible solutions. However, here we look at the problem more carefully keeping the symmetry of the growth process in mind. From the Plancherel measure (\ref{eq:Plancherelmeasure2}) we see that at any level $k$, two Young diagrams related to each other by transposition, have same probability $\cP(\lambda_k)$. Therefore the large $N$ solution of (\ref{eq:saddeq2}) must be invariant under transposition. Young diagrams, symmetric under transposition, are called rectangular diagrams \cite{Kerov1,Hora:2237389}.

\begin{figure}[h]
	\begin{center}
		\includegraphics[width=5cm,height=5cm]{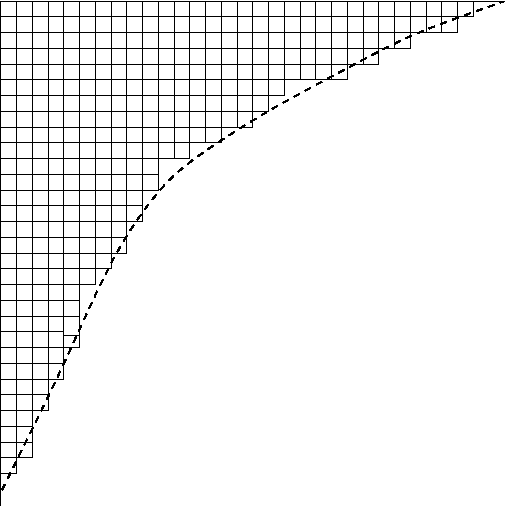}
		\caption{A Young diagram in English notation for automodel class ($0<\xi<1/2$).}
		\label{fig:YD-english-automodel}
	\end{center}
\end{figure}

Following \cite{duttagopakumar}, we can take the following ansatz for $u(h)$ to get a rectangular Young diagram 
\ben \label{eq:uhsym}
\begin{split}\label{eq:sym-ansatz}
	u(h) = \bigg\{ {1 \qquad \hspace{.16cm}  h\in [0,p) \atop
	 \tilde u(h) \quad h\in (p,q].}
\end{split}
\een
To solve the saddle point equation we define a resolvent
\ben\label{eq:Hresolvent}
H(h) = \int_{h_L}^{h_U} \frac{u(h')dh'}{h-h'}.
\een
After a little algebra, we find that the resolvent $H(h)$ is given by \cite{duttagopakumar}
\ben \label{eq:resolventsym}
H(z)= \ln\lB\frac{h \left(h-1 -\sqrt{(h-1)^2-4 \xi ^2}\right)}{2 \xi ^2}\rB.
\een
The resolvent is same as the moment generating function for the rectangular diagrams defined in (\ref{eq:momgenfunc}) \cite{Kerov1}. The resolvent has a branch cut in the complex $h$ plane. Young diagram density is given by the discontinuity of $H(h)$ about the branch cut
\be\label{eq:sym-answer}
\tilde u(h) = \frac1\pi \cos^{-1} \lB  \frac{h-1}{2\xi}\rB, \quad \for\quad p\leq h\leq q.
\ee
The supports $p$ and $q$ are given by,
\be
p=1-2\xi, \quad q=1+2\xi.
\ee
This particular class of solution exists subject to the following condition
\be
k'=\xi^2.
\ee
Since $p\geq 0$, this solution is valid for $0\leq\xi\leq 1/2$. From the definition of $\xi$ ($\xi^2 = z k'$) we also see that this solution exists for 
\ben
\text{either} \quad \xi =k'=0 \quad \text{or} \quad z=1.
\een
The case $\xi=k'=0$ is trivial. This means there is no box in the Young diagram. The non-trivial solution corresponds to $z=1$ (i.e. fugacity is one and hence zero chemical potential). It is easy to check that the Young diagram is invariant under transposition. The height of the first column can be calculated from equation (\ref{eq:hook2}) and is given by $2\xi$ which is similar to the length of the first row. Also the function $\tilde u(h)$ is symmetric about body diagonal. A typical Young diagram for this class is depicted in figure \ref{fig:YD-english-automodel}.
In this phase the renormalised number of boxes (i.e. $k'$) in a Young diagram grows from $k'=0$ to $k'= 1/4$ as $\xi$ changes from $0$ to $1/2$. For any value of $\xi$ between $0$ and $1/2$ the dominant Young diagram is always symmetric under transposition and hence a rectangular diagram. The limiting value $\xi=1/2$ (GWW transition point) corresponds to the distribution
\be\label{eq:terminaluh}
\tilde{u}(h) =\frac1\pi \cos^{-1}(h-1).
\ee
This terminal distribution is same as the universal curve or the limit shape obtained by \cite{LogShe,VerKer77}. Hence we see that the limit shape Young diagram corresponds to GWW transition point in matrix model side. 

We calculate Plancherel measure (\ref{eq:Plancherelmeasure2}) for this dominant diagram. Following \cite{duttagopakumar}, the Plancherel measure  in large $N$ limit is given by
\be
\frac1{N^2}\ln \cP_{\lambda_k} = \int_{0}^{q} dh^{} u(h)  \ \Xint-_0^q dh' u(h') \ln|h-h'| -2 \int_0^q u(h) h \ln h \ dh  +k'+1+k'\ln k'.
\ee
Evaluating the right hand side for symmetric solutions (\ref{eq:sym-ansatz}) and (\ref{eq:sym-answer}) we get
\be
\cP_{\lambda_k} = 1 +O\lb 1/N\rb.
\ee

Thus we see that in the large $k$ (or large $N$) limit the symmetric solution (\ref{eq:sym-ansatz}, \ref{eq:sym-answer}) is the maximum probable solution. Probability of having other diagrams is suppressed by powers of $N$. This gives an alternate proof of \emph{limit shape} theorem of Vershik-Kerov and Logan-Shepp result \cite{VerKer77,LogShe}.

The partition function (\ref{eq:GCPF3}) admits another phase. The dominant diagrams in that phase are not self-transpose. See \ref{app:asymetricsol} for details. Growth of such diagrams are different than that studied in section \ref{sec:plancherel}.

\subsection{A connection with automodel}

To make a precise connection between automodel class and the above solution, we need to set up a dictionary between the variables defined in $(u,v)$ plane and $(h,x)$ plane. The relation between \emph{French} notation and \emph{English} notation is $Y=n\ \tand \ X=x$. We use the following transformation between $(n,x)$ and $(u,v)$ so that 
$u=v=2$ point is mapped to $n=1, x=0$
\ben
\begin{split}
	\frac u2 &= n-x \\
	\frac v2 &= n+x.
\end{split}
\een
Using this mapping one can show that the Young diagram distribution function (\ref{eq:uhdef}) is related to $v'(u)$ in the following way
\be
\label{eq:Ivanov-our-map}
u(h) = \frac12 -\frac12 v'(u) \quad \with \quad u=2(h-1).
\ee 
One can also check that the terminal diagram (\ref{eq:terminaluh}) is exactly same as the limit shape defined in (\ref{eq:kerovfunction}). Thus we see that the Young diagram density $u(h)$ is related to \emph{charge} $\sigma(u)$ defined in (\ref{eq:youngcharge}) by $u(h) = - \sigma'(u)$. The resolvent (\ref{eq:resolventsym}) for this symmetric solution is same as the moment generating function for charges (\ref{eq:momgenfunc}).

We also observe that the symmetric distributions (\ref{eq:uhsym}) for $0<\xi<1/2$ satisfies,
\be\label{eq:automodeluh}
\dow_\xi u(h,\xi) +\frac{h-1}{\xi} \dow_hu(h,\xi)=0.
\ee
Since for this branch we have $\xi^2=k'$, the above equation can be written as,
\be\label{eq:automodeluh2}
\dow_{k'} u(h,k')+\frac{h-1}{2k'} \dow_h u(h,k') =0.
\ee
Thus we see that the Young diagram density satisfies the automodel equation\footnote{Please note the difference in notation. Here $u$ is Young diagram density and $h$ is hook number.} (\ref{eq:automodeleqnsigma}) with $k'$ playing the role of automodel time $t$. This is natural to expect that the renormalised box number $k'$ plays the role of growth parameter $t$ in Kerov's paper \cite{Kerov1}. Hence we conclude that the no-gap phase of GWW model (\ref{eq:cZ}) or  (\ref{eq:cZ2}), in the limit of large box number, captures the Plancherel growth of Young diagrams and the dominating Young diagrams belong to the automodel class of Kerov.

\section{The Hilbert space for Plancherel growth}
\label{sec:hilbertspace}

As we have reviewed in section \ref{sec:umm}, the unitary matrix models admit a phase space or droplet description in two dimensions spanned by $(h,\theta)$. Following \cite{Chattopadhyay:2020rle} we show that the growth of Young diagrams can be described in terms of evolution of \emph{classical} or coherent states in a Hilbert space. The goal of this section is to construct the Hilbert space by quantising the free Fermi droplets.

\subsection{Droplet quantisation}\label{sec:fluctuation}

The single particle Hamiltonian (\ref{eq:singleh}) is not in diagonal form. We define a new variable
\ben
\bar h = h-S(\theta).
\een
In terms of this new variables $(\bar h, \theta)$ the Hamiltonian density is given by
\ben\label{eq:singleh2}
\mathfrak{h}(\bar h,\theta) = \frac{\bar h^2}2 -\frac{\pi^2}2 \rho(\theta)^2
\een
and the droplet boundary is given by
\ben\label{eq:boundaryeqn2}
\bar h_{\pm}(\theta) = \pm \pi \rho(\theta).
\een
The phase space Hamiltonian (\ref{eq:Hh}) can be calculated by integrating over $\bar h$ from $\bar h_{-}(\theta)$ to $\bar h_{+}(\theta)$ and is given by,
\ben\label{eq:Hh1}
H_h = \frac1{3\pi \hbar}\int d\theta \ \pi^3 \rho(\theta)^3.
\een
The phase space boundary is doubly degenerate, i.e. for a given $\theta$ there are two boundaries, $\bar h_{\pm}$ corresponding to two signs on the right hand side of (\ref{eq:boundaryeqn2}). Hence, we write the Hamiltonian (\ref{eq:Hh1}) over the two segments of droplets \cite{Maoz:2005nk}
\ben\label{eq:Hhtwoboundary}
H_h = \frac1{6\pi \hbar}\left[\int_+ d\theta \ \bar h^3_+(t,\theta) - \int_- d\theta \ \bar h^3_-(t,\theta)  \right].
\een
From the single particle Hamiltonian (\ref{eq:singleh2}) we find that the equations of motion satisfied by the droplet boundaries are given by\footnote{Equations of motion from (\ref{eq:singleh2}) are given by %
\ben\label{eq:eom}
\dot{\bar h}(t) = -\frac{\partial \mathfrak h}{\partial \theta} = \pi^2 \rho(\theta) \rho'(\theta), \qquad \dot \theta(t) = \frac{\partial \mathfrak h}{\partial \bar h} = \bar h(t)
\een
Substituting the boundary relations (\ref{eq:boundaryeqn2}) in these equations, one finds (\ref{eq:boundeq}).},
\ben\label{eq:boundeq}
\dot{\bar h}_{\pm}(t,\theta) = \bar h_\pm(t,\theta)\bar h_\pm'(t,\theta)
\een
{where the dot and prime denote derivative with respect to $t$ and $\theta$ respectively} \footnote{{At this point one should note that (\ref{eq:boundeq}) is in the standard from of an inviscid Burger's equation, which is a KdV type equation. The appearance of KdV type equations is quite common in random matrix theories which indicates that the problem is exactly solvable. For the Fermi fluid picture that appears in the context of unitary matrix model one can actually construct an infinite set of commuting \emph{Hamiltonians/conserved charges} \cite{Avan:1991kq,Avan:1991ik}, signifying the integrability of the system. In fact, in \cite{Avan:1991ik} the authors have used analogous variables of (\ref{eq:hpm1}) to deduce the Lax pairs with the same Hamiltonian as (\ref{eq:Hhtwoboundary}). Further, the commuting Hamiltonians described in \cite{Avan:1991ik,Jevicki:1993qn} can be shown to be similar to different moments of the diagram (\ref{eq:moments}) and the resolvent (\ref{eq:Hresolvent}) or similarly the moment generating function (\ref{eq:momgenfunc}) includes information about the full \emph{spectrum} of the Hamiltonians. Therefore from the integrability structure emergent from the matrix model side one can show that all the moments of the diagram as described by \cite{Kerov1} is conserved. In the current context we are interested in \emph{canonically quantising} the droplets, rather than looking at the underlying integrability structure of (\ref{eq:Hhtwoboundary}).}}.

In order to quantise the droplets we find a simplectic form such that the Hamilton's equation
\be\label{eq:Hameq2}
\dot {\bar h}_{\pm}(t,\theta) = \{\bar h_\pm(t,\theta),H_h\}
\ee
would coincide with (\ref{eq:boundeq}) \cite{Grant:2005qc,Maoz:2005nk}. To achieve the above goal we introduce Poisson brackets between $\bar h_{\pm}(t,\theta)$ and $\bar h_{\pm}(t,\theta')$
\ben\label{eq:poisson}
\{\bar h_\pm(t,\theta),\bar h_\pm(t,\theta')\} = \pm {\pi \hbar} \delta'(\theta-\theta')\quad \text{and} \quad \{\bar h_+(t,\theta),\bar h_-(t,\theta')\} =0.
\een
It is easy to check that using these Poisson brackets, equation (\ref{eq:Hameq2}) boils down to (\ref{eq:boundeq}). {Also,
\be
\{ H_h, \cA \}=0
\ee
where
\be
\mathcal{A}=\int_{-\pi}^{\pi} d\theta\, (\overline{h}_+(t,\theta) - \overline{h}_-(t,\theta))
\ee
is the area of the phase space droplets. This implies that the area is preserved under classical time evolution.}

The circular droplet ($\rho(\theta) =\frac1{2\pi}$) corresponds to $\bar h_\pm(t,\theta)=\pm \frac12$ and satisfy the classical equations of motion (\ref{eq:boundeq}). We consider quantum fluctuations about this classical circular droplet
\ben\label{eq:hpmode}
\bar h_{\pm}(t,\theta) = \pm\frac12+ \hbar\ \tilde h_\pm(t,\theta).
\een
We also assume that the fluctuations preserve the total area of the droplet. This implies that 
\ben\label{eq:tildehcons}
\int_{-\pi}^{\pi} d\theta (\tilde h_+(t,\theta) - \tilde h_-(t,\theta))=0.
\een
The Hamiltonian (\ref{eq:Hhtwoboundary}) for these fluctuations is given by
\ben\label{eq:Hhtwoboundary2}
H_h = \frac{1}{12\hbar} + \frac{\hbar}{4\pi}\int_{-\pi}^{\pi} \lb \tilde h_+^2(t,\theta) + \tilde h_-^2(t,\theta) \rb d\theta + \frac{\hbar^2}{6 \pi} \int_{-\pi}^{\pi} \lb \tilde h_+^3(t,\theta) - \tilde h_-^3(t,\theta) \rb d\theta.
\een

From equation (\ref{eq:poisson}), we find that the Poisson bracket for $\tilde h_\pm(t,\theta)$ are given by
\ben\label{eq:poisson2}
\{\tilde h_\pm(t,\theta), \tilde h_\pm(t,\theta')\} =\pm \frac{\pi}{\hbar} \delta'(\theta-\theta') \quad \text{and} \quad \{\tilde h_+(t,\theta),\tilde h_-(t,\theta')\} =0.
\een
To quantise the above classical system we promote the Poisson brackets (\ref{eq:poisson2}) to commutation relations
\be\label{eq:commutation}
\left[ \tilde h_{\pm}(t,\theta), \tilde h_{\pm}(t,\theta')\right] = \pm \pi i \delta'(\theta-\theta')\quad \text{and}\quad [\tilde h_{+}(t,\theta),\tilde h_-(t,\theta')]=0.
\ee
We decompose $\tilde{h}_\pm(t,\theta)$ into Fourier modes
\ben\label{eq:hpmode}
\tilde h_{+}(t,\theta) = \sum_{n=-\infty}^\infty a_{-n}(t) e^{i n \theta}
\een
and
\ben\label{eq:hmmode}
\tilde h_{-}(t,\theta) =  - \sum_{n=-\infty}^\infty b_{n}(t) e^{i n \theta}.
\een
From the constraint equation (\ref{eq:tildehcons}), we see that the zero-modes $a_0$ and $b_0$ are equal up to a sign
\ben
a_0=-b_0=\pi_0.
\een
It follows from equation (\ref{eq:poisson2}) that the Fourier modes $a_n$ and $b_n$ satisfy $U(1)$ \emph{Kac-Moody} algebra
\ben \label{eq:U1KM}
[a_m(t),a_n(t)]={\frac{1}{2}}m\delta_{m+n},\quad [b_m(t),b_n(t)]={\frac{1}{2}}m\delta_{m+n}, \quad \tand \quad [a_m(t),b_n(t)]=0.
\een
Since $\tilde h_\pm(t,\theta)$ are real we have $a_{-n}=a_n^\dagger$ and $b_{-n}=b_n^\dagger$. The Hamiltonian (\ref{eq:Hhtwoboundary2}) in terms of these modes is given by (up to over all constant factors)
\ben
\tilde H = H_+ + H_-
\een
where
\ben
\begin{split}
H_{+} = & \frac{\hbar}{2}a_{0}^2 + \frac{\hbar^2}{3}a_{0}^3 - \frac{\hbar^2}{24}a_{0} + \hbar (1+2\hbar a_0)\sum_{n>0} a_n^\dagger a_n \\
 &+ \hbar^2\sum_{m,n>0} \lb a_{m+n}^\dagger a_m a_n+ h.c. \rb
\end{split}
\een
and $H_-$ has a similar expression in terms of modes $b_n$'s.\\
Since $\pi_0$ commutes with all the $a_n, b_n$ and hence with $\tilde H$, application of $a_n$'s and $b_n$'s cannot change the eigenvalue of $\pi_0$. Therefore the Hilbert space is constructed upon a one parameter family of vacua $\ket{s,s}\equiv \ket s$ where,
\ben
\label{eq:KMprimary}
\begin{split}
   & a_n\ket{s} = 0, \quad b_n\ket{s} = 0 \quad \for\  n>0\\
\tand \quad & a_0\ket{s} = - b_0\ket{s} =\pi_0\ket{s} = s \ket{s}.
\end{split}
\een
Starting with the primary $\ket{s}$ we can now construct a Hilbert space $\cH$ which is a $s$ charged module. Let us denote $\cH_+$ and $\cH_-$ as Hilbert spaces associated with $a$ and $b$ sectors respectively. The commutativity of $a$ and $b$ operators implies that the full Hilbert space is factorizable into $\cH_+$ and $\cH_-$ i.e. $\cH=\cH_+ \otimes \cH_-$.

A generic excited state in $\cH$ is given by
\ben\label{eq:basisstatek}
\ket{\vec k,\vec l} = \prod_{n,m=1}^{\infty} \lb a^\dagger_n\rb^{k_n}\lb b^\dagger_m\rb^{l_m}\ket{s}.
\een
The $\vec k$ and $\vec l$ vectors correspond to excitations in the upper and lower Fermi surfaces. Since $a_n$ and $b_n$ commute, generic excitation $\ket{\vec k, \vec l} \in \cH$ can be written as a direct product of states belonging to the two sectors i.e. $\ket{\vec k, \vec l}=\ket{\vec k} \otimes \ket{\vec l}$.

The excited states in $\cH_+$ are orthogonal with the normalization
\ben\label{eq:knormalization}
\langle {\vec{k'}}|{\vec k}\rangle =z_{\vec k}\delta_{\vec k \vec{k'}}\ \ \text{where}\ z_{\vec{k}}=\prod_j k_j! {\left(\frac{j}{2}\right)}^{k_j}
\een
and has $\pi_0$ eigenvalue $s$. They also satisfy the completeness relation
\ben
\sum_{\vec k}\frac{1}{z_{\vec k}} \ket{\vec k}\bra{\vec k} = {\mathbb{I}}_{\cH_+}
\een
and hence form a basis in $\cH_+$. The particle like excited states $\ket{\vec k}$ in either sectors are eigenstates of the free Hamiltonian of the corresponding sectors
\ben
H^\pm_{\text{free}}\ket{\vec k} = \hbar(1 \pm 2s\hbar)\lb\sum_{n=1}^\infty n k_n\rb \ket{\vec k},
\een
but not an eigenstate of the full Hamiltonian. The interaction Hamiltonian is such that its action on an excited state doesn't change the level of the state. Explicitly stated, the interaction Hamiltonian takes a state $\ket{\vec k}$ to $\ket{\vec{k'}}$ satisfying $\sum_n n k_n =\sum_n n{k'}_n$. The expectation value of $\bar h_+(t,\theta)$ operator in $\ket{\vec k}$ state is $(1/2+\hbar s)$. The quantum dispersion, $\Delta \bar{h}_+$ in $\ket{\vec k}$ state is $O(\hbar)$ and hence a generic $\ket{\vec k}$ state can be interpreted as quantum fluctuations over the ground state.

A generic coherent state in $\cH_+$ can be defined as\footnote{{A rather interesting feature of coherent states written in this form is pointed out by \cite{Dijkgraaf:1992kv, Segal:1985aga, Date:1981qy}. One can show that the inner product between the coherent state (\ref{eq:coherentstate}) and a state lying in the Grassmannian is tau function of the KP hiererchy.}}
\ben\label{eq:coherentstate}
\ket{\t_+} = \exp\lb {\sum_{n=1}^\infty \frac{2\t_{n}^+ a_n^\dagger}{n\hbar}}\rb \ket{s}.
\een
The state $\ket{\t_+}$ is not normalised. The normalization is given by 
\ben
\braket{\tau_+^a}{\tau^b_+} = \exp\lb\sum_n \frac{4\t_{n+}^a\t_{n+}^b}{n\hbar^2}\rb.
\een
The coherent state $\ket{\t_+} $ is an eigenstate of $a_n$ ($\forall\, n>0$) operator with eigenvalue $\t_{n}^+/\hbar$. A coherent state $\ket{\t_+}$ can be expanded in $\ket{\vec k}$ basis in the following way
\ben
\ket {\t_+} = \sum_{\vec k} \frac{\t^+_{\vec k}}{z_{\vec k}}\ket{\vec k}, \quad \where \quad \t^+_{\vec k} = \prod_m \lb\frac{\t^+_{m}}{\hbar}\rb^{k_m}.
\een
The expectation value of $ \bar{h}_+$ operator in a coherent state $\ket {\t_+}$ is given by
\ben
\omega_{\t_+}(z)= \frac{ \bra{\t_+}\frac{\bar h_+(z)}{\pi} \ket{\t_+}}{\braket{\t_+}{\t_+}} =\frac{1}{2\pi} + \frac{s\hbar}{\pi}+\frac{1}{\pi}\sum_{n>0}\t^+_{n}\lb z^n +\frac{1}{z^n}\rb.
\een
Value of $\omega_{\t_+}(z)$ on the unit circle ($|z|=1$) in the complex $z$ plane is given by, 
\ben\label{eq:coherentprofile}
\begin{split}
\omega_{\t_+}(\theta) \equiv \omega_{\t_+}(z=e^{i\theta}) & = \frac{1}{2\pi}  + \frac{s\hbar}{\pi} + \tilde{\omega}_{\t_+}(\theta)\\
\text{where} \quad \tilde{\omega}_{\t_+}(\theta) &=\frac{2}{\pi} 
\sum_{n>0}\t^+_{n} \cos n\theta.
\end{split}
\een
Since the quantum dispersion of $\bar{h}_+$ in a coherent state is zero, we call such states \emph{classical}. 

We set up a map between a state $\ket{\psi} \in \cH_+$ and shape of the upper Fermi surface
\ben\label{eq:maping}
\ket{\psi} \ra \{\bra{\psi}\bar{h}_+(\theta)\ket{\psi}\}.
\een
This maps the following three types of states in $\cH_+$ to three different types of shapes of the upper Fermi surface $\bar h_+$. Such a mapping has been also discussed in \cite{Chattopadhyay:2020rle}. Expectation value of $\bar h_+$ in ground state is given by $\frac12 + s \hbar$ and has zero dispersion. Thus, ground state $\ket s$ corresponds to an overall shift of $\bar h_+$ by an amount $s/N$ over the classical value. Generic \emph{normalised} excited states correspond to ${\cal O}(\hbar)$ ripples on $\bar h_+$. In this case, the expectation value of $\bar{h}_+$ picks up a non-zero dispersion at ${\cal O}(\hbar)$. ${\cal O}(1)$ deformation to the upper Fermi surface $\bar{h}_+$ is mapped to coherent states $\ket{\tau_+}$ in $\cH_+$ as is evident from \ref{eq:coherentprofile}. Here we have assumed $\langle \tau_+|\bar{h}_+(\theta)|\tau_+\rangle$ is a single valued function of $\theta$. An important thing to note here is that the classical deformations do not disturb the quadratic profile of the droplets \emph{i.e} for a given $\theta$, there exists unique values of $h_\pm(\theta)$.

We now consider the mapping between the automodel diagrams and states in the Hilbert space. The automodel diagrams corresponds to
\ben
\bar h_\pm(\theta) = \pm \lb \frac12 +\xi \cos\theta\rb. 
\een
The set of coherent states with $\tau_1=\xi/2$ , $\tau_n = 0$ $\forall n > 1$ are mapped to the automodel diagrams. In particular, the state corresponding to the limit shape is given by
\ben
|\tau* \rangle = \text{exp}\left(\frac{a_{1}^{\dagger}\xi}{\hbar}\right) |0\rangle. \quad
\een
Thus Plancherel growth of Young diagrams is mapped to evolution of classical states \cite{Chattopadhyay:2020rle}. One can also compute the transition amplitude of such evolution process. The ground state $\xi=0$ is an eigenstate of the Hamiltonian and hence its evolution is trivial. Therefore, we take our initial state as
\ben
\ket{\tau_i} = \exp{2a_1^\dagger}\ket{s}.
\een
Such state corresponds to a droplet with $\frac12 +\hbar \cos\theta$. The transition amplitude for evolution from this state to a automodel state in time $T$ is therefore given by \cite{Chattopadhyay:2020rle}
\ben
\mathbb{T} \sim \sum_R\lb \frac{\dim R}{k!}\rb^2 \lb \frac{\xi}{\hbar}\rb^{|R|} e^{i T \hbar C_2(R)}
\een
where $C_2(R)$ is given by
\ben
C_{2}(R) = N \sum_{i=1} l_{i} + \sum_{i=1}l_i(l_i - 2 i + 1)
\een
Here $l_i$ is the number of boxes in the $i$th row of the Young diagram $R$.

Excitations over this state correspond to $O(\hbar)$ fluctuations of the limit shape. Consider the states
\ben\label{state:coh}
\text{exp}\left( \frac{a_{1}^{\dagger}\xi}{\hbar} + \sum_{n>1}\frac{2\alpha_{n} a_{n}^{\dagger}}{\sqrt{n}} \right) |0\rangle
\een
The expectation value of $\bar h_+(\theta)$ in such a state gives
\ben\label{expect:coh}
\frac{1}{2} + \xi \cos{\theta} + 2\hbar \sum_{n>1}\sqrt{n}\alpha_{n}\cos{n\theta}
\een
which is essentially the result (\ref{eq:flucLS}) of Ivanov and Olshanki \cite{2003math.4010I}.

\section{Conclusion}\label{sec:conclusion}

In this paper we show that the growth of Young diagrams equipped with Plancherel measure can be studied through a simple unitary matrix model, namely Gross-Witten-Wadia model. Considering the growth process to be Markovian and governed by Plancherel probability naturally we write down a partition function for such growth process. Plancherel growth also serves as an interesting toy model to study growth or melting of \emph{2d} crystals \cite{Eynard:2009nd,Eynard:2008mt}. Strictly speaking, the partition function can be evaluated exactly as one can explicitly perform the sum over all possible representations. However, in order to capture the growth process via saddle point analysis in the limit of large box number, we introduce a cut-off $N$ on the number of rows in the Young diagrams. The regularised partition function turns out to be identical to that of the GWW model with $N$ playing the role of the rank of the gauge group. We performed a saddle point analysis of the regularised partition function and find that in the continuum limit the partition function is dominated by a class of diagrams. The class is isomorphic to no-gap phase of GWW model. Different diagrams in this class are characterised by a single parameter $0<\xi<1/2$. The limiting diagram $\xi \ra 1/2$ corresponds to limit shape and mapped to GWW transition point in the eigenvalue side. This gives an alternate proof of the limit shape theorem of large Young diagrams \cite{LogShe,VerKer77} . In \cite{Kerov1}, Kerov introduced a differential model for the growth of Young diagrams, known as the \emph{automodel}. We find that in continuum limit the above one parameter class of diagrams falls in the automodel class of Kerov with renormalised box number playing the role of \emph{time}.

There exists a correspondence between unitary matrix models and free Fermi droplets in two dimensions. Using this correspondence, we see that the evolution of Young diagrams in automodel class can be mapped to different shapes of incompressible fluid droplets in two dimensions. Automodel evolution corresponds to area preserving deformation of these fluid droplets. Such identification was possible due to the equivalence between GWW model and automodel partition function \cite{duttagopakumar}. Since eigenvalues of unitary matrices behave like position of free fermions, two dimensional fluid droplets are identified with classical phase space of these free fermions \cite{duttagopakumar,Chattopadhyay}. Although the model under consideration is non-dynamical, we introduce time in an \emph{ad-hoc} manner by identifying our eigenvalue distribution with the Fermi distribution at zero temperature thus constructing a one particle Hamiltonian. We subsequently use the techniques developed in \cite{Chattopadhyay:2020rle} to provide a Hilbert space description for the growth process of Young diagrams. In order to achieve this, we quantise these free Fermi droplets and find that the boundary modes satisfy Kac-Moody algebra. Edge excitations of fractional quantum Hall fluid also satisfy similar algebra \cite{Wen:1995qn}. We construct the Hilbert space associated with the quantised droplets and identify a mapping between Young diagrams and coherent states in the Hilbert space. Firstly, the ground state of the Hilbert space corresponds to the circular shape of the Fermi surfaces and are identified with no box diagram. Secondly, a generic excited state corresponds to $\mathcal{O}(1/N)$ ripples on the Fermi surface. Thirdly, a $\mathcal{O}(1)$ deformation to the Fermi surfaces are captured by coherent states in the Hilbert space and is mapped to regularised large $N$ diagrams. Certain coherent states in the Hilbert space correspond to automodel Young diagrams. Young diagrams in automodel class are characterised by a one parameter class of coherent states. Hence, growth of Young diagrams are mapped with evolution of these coherent states in the Hilbert space. We also observe that the shape of arbitrary fluctuations matches with the fluctuations of limit shape studied earlier by Ivanov and Olshanki. \cite{2003math.4010I}.

\vspace{1cm}

\noindent{\bf Acknowledgments:} We are grateful to Rajesh Gopakumar, Debashis Ghoshal, Dileep Jatkar, Parikshit Dutta, Nabamita Banerjee, Ashoke Sen for helpful and illuminating discussions. The work of AC is supported by a Simons Foundation Grant Award ID 509116 and by the South African Research Chairs initiative of the Department of Science and Technology and the National Research Foundation. The work of SD is supported by the grant no. EMR/2016/006294 and MTR/2019/000390 from the SERB, Government of India. SD also acknowledges the Simons Associateship of the Abdus Salam ICTP, Trieste, Italy. AC and SD would like to thank the hospitality of ICTP where part of the project was completed. We all indebted to people of India for their unconditional support towards research in basic science.

\appendix

\section{Analysing Unitary Matrix Models}\label{app:ummlysis}
\subsection{Eigenvalue analysis}
\label{app:evanalysis}
In this section, we will briefly review the eigenvalue analysis, the Young diagram analysis and their connection in the context of the so-called single plaquette model. Interested readers may refer to \cite{Chattopadhyay} for further details.

In the diagonal basis the partition function of the single plaquette model described by the action \eqref{eq:partfuncplaq} is given by (up to an overall volume dependent factor),
\ben\label{eq:Zevbasis}
Z = \int \prod_{i=1}^N d\theta_i \prod_{i<j} \sin^2\lb\frac{\theta_i-\theta_j}{2} \rb \exp{N\sum_n \frac{2\beta_n}{n}\sum_{i=1}^N\cos n\theta_i}.
\een
In the large $N$ limit, the discrete eigenvalues $\theta_i$ goes over to a smooth continuous function $\theta(x)$ where $x \in [0,1]$. One can write down a saddle point equation in the continuum limit which is given by
\ben\label{eq:evsaddle}
{\Xint -}_0^1 dy \cot\lb \frac{\theta(x)-\theta(y)}{2}\rb = \sum_n \beta_n \cos n\theta(x).
\een
Defining a distribution function (called eigenvalue density) $\rho(\theta)$ for eigenvalues over a unit circle
\ben
\rho(\theta) = \frac1N \sum_{i=1}^N\delta(\theta-\theta_i) = \frac{dx}{d\theta}.
\een
the saddle point equation (\ref{eq:evsaddle}) can be recast as
\ben
\Xint- d\theta' \cot\lb \frac{\theta-\theta'}{2}\rb \rho(\theta') = \sum_n \beta_n \cos n\theta, \quad \with \quad \int d\theta \rho(\theta) = 1.
\een

\subsection{Young diagram analysis}
\label{app:YDalysis}

The exponential appearing in the partition function of the single plaquette model can be expanded into an infinite series which runs over representations $R$ of unitary group $U(N)$,
\ben \label{eq:ZplaqsumoverR}
Z = \sum_R \sum_{\vec k} \frac{\varepsilon(\vec \b,\vec k)}
{z_{\vec k}} \sum_{\vec l} \frac{\varepsilon(\vec \b,\vec l)} {z_{\vec
		l}} \chi_{R}(C(\vec{k}))\chi_{R}(C(\vec{l})).  
\een
Here $\chi_{R}(C(\vec{k}))$ is the character of conjugacy class
$C(\vec{k})$ of permutation group $S_{K}$, $K=\sum_n n k_n$ and
\ben
\varepsilon(\vec \b, \vec k) =
\prod_{n=1}^{\infty}N^{k_n}\b_{n}^{k_n}, \quad
z_{\vec k} = \prod_{n=1}^{\infty} k_{n}! n^{k_n}.
\een
Sum of representations of $SU(N)$ can be written as a sum over different
Young diagrams. Say, $n_i$ is the number of boxes in the $i$-th row of the Young diagram. We decompose the sum over representations as a sum over all possible Young diagrams that one can draw for the symmetric group. We introduce $N$ new variables which are defined as:
\begin{eqnarray}\label{eq:h-nrelation}
h_i = n_i + N -i \qquad \forall \quad i =1, \cdots, N.
\end{eqnarray}
$h_i$'s are called hook numbers and they satisfy the following constraint
\be
\label{hmonotonicity} h_1> h_2> \cdots > h_N \geq 0.  
\ee
In terms of the new variables $h_i$, the partition function (\ref{eq:ZplaqsumoverR}) is therefore given by,
\ben 
Z=\sum_{\{h_i\}} \sum_{\vec k, \vec l}
\frac{\varepsilon(\vec \b, \vec k) \varepsilon(\vec \b, \vec
	l)}{z_{\vec k} z_{\vec l}} \chi_{\vec h}(C(\vec k)) \chi_{\vec
	h}(C(\vec l)) 
\een
where $\vec h$ denotes a particular representation.

As before we define continuous variables in the large $N$ limit
\begin{eqnarray}\label{eq:contvardef}
{h_{i}\over N}=h(x), 
\quad\quad k_{n}=N^{2}k'_n,
\quad\quad
x={i\over N} \quad \with \quad x\in [0,1].
\end{eqnarray}
Writing characters $\chi_{\vec h}$ in terms of $h(x)$ and
$k_n'$, partition function (\ref{eq:ZplaqsumoverR}), in large $N$ limit, can
be written as
\be\label{eq:pfyt1}
Z = \int [d h(x)] \prod_n\int dk_n' dl_n' \exp\lB -N^2 
S_{{eff}}^h [h(x),\vec {k_n'},\vec{l_n'}]\rB,
\ee
where $S^h_{{eff}}$ is the effective action in Young diagram basis. Dominant contribution to partition function comes from those representations which minimise the effective action $S^h_{{eff}}$. To find the most dominant
representations at large $N$, we introduce a function called Young
tableaux density defined as,
\be
u(h) = - {d x\over d h}.
\ee

Thus we see that the unitary matrix model under consideration can be expressed in two different basis - eigenvalue basis (picture I) and Young diagram basis (picture II). Different large $N$ phases can be characterised either by distributions of eigenvalues or Young diagrams. Therefore it is expected that there is a relation between these two pictures. We now discuss this relation.

\subsection{Finding the connection between two pictures}
\label{app:connection}

The character of permutation group is given by Frobenius formula
\ben\label{eq:frobenius}
\chi_R(C(\vec k)) =\left[ \Delta(x) \prod_n \left(P_n(x)\right)^{k_n}\right]_{(h_1,h_2,\cdots h_N)}
\een
where
\ben\label{eq:Pn}
P_n(x) = \sum_{i=1}^N x_i^{n}, \quad \Delta(x) =\prod_{i<j}(x_i-x_j)
\een
and $(x_1,\cdots,x_N)$ are set of auxiliary variables. The notation
$[\cdots]_{(h_{1},\cdots, h_{N})}$ implies
\begin{equation}
\lB f(x)\rB_{(h_{1},\cdots, h_{N})}=\textrm{coefficient of }
x_{1}^{h_{1}}\cdots x_{N}^{h_{N}}\textrm{ in }f(x) .
\end{equation}
Promoting $N$ auxiliary variables to $N$ complex variables $(z_1,\cdots,z_N)$,  Frobenius formula \eqref{eq:frobenius} can be written as a residue of a complex function as follows
\ben\label{eq:character1}
\chi[h(x), C(\vec{k'})]= \lb\frac1{2\pi i}\rb^N 
\oint {[\cD z(x)]\over z(x)}
\exp[-N^2 S_{\chi}(h[x],\vec{k'})]\ .
\een
This is subsequently used in \eqref{eq:ZplaqsumoverR} to obtain
\ben\label{eq:pftotal}
Z = \prod_n \int dk_n' dl_n' \int  [dh(x)]\oint  [dz(x)]
\oint [dw(x)] \int_{-\I}^\I dt \int_{-\I}^\I ds
\exp\lb - N^2 S_{{total}}\lB h,z,w,\vec{k'},
\vec{l'},t,s\rB \rb\nonumber \hspace*{-1cm}\\
\een
where,
\begin{eqnarray}\label{eq:effStotal}
\begin{split}
- S_{{total}}\lB h,z,w,\vec{k'},
\vec{l'},t,s\rB &= \sum_n \bigg[ k_n'\lb1+ \lna{{\b_n Z_n\over n k_n'}}\rb 
+l_n'\lb1+ \lna{{\b_n W_n\over n l_n'}}\rb \bigg]\\
&\qquad +\frac12 \int_0^1 dx \Xint-_0^1 dy\lb \ln|z(x)-z(y)|
+\ln|w(x)-w(y)|\rb\\
& \qquad -\int_0^1 dx h(x) \ln z(x) w(x) 
+i t \lb \sum_n n k_n' -K'\rb +i s \lb \sum_n n l_n' -K'\rb.
\end{split}
\end{eqnarray}
Extremising $S_{total}$ with respect to $h(x)$, $z(x)$, $w(x)$, $k_n'$ and $l_n'$, we see that the saddle point equations are given by
\ben\label{eq:saddlepoint}
\sum_n {n k_n'\over Z_n} z^n(x) + \Xint-_0^1 dy
{z(x)\over z(x)-z(y)}
-h(x) =0 \qquad \textrm{(variation w.r.t $z(x)$)}
\een
where,
\be
Z_n = \int_0^1 dx z^n(x).
\ee
Similar equation can be obtained when we vary action with respect to $w(x)$. Variation with respect to $h(x)$ gives
\be\label{eq:saddleh}
z(x)w(x) = e^{-i(t+s)} = \text{constant (independent of} \ x).
\ee
Finally, variation with respect to $k_n'$ and $l_n'$ provides,
\ben\label{eq:saddlekl}
{\b_n Z_n\over n k_n'} = e^{-i t n}, \quad \text{and}
\quad {\b_n W_n \over n l_n'} = e^{-i s n}.
\een
Since the contours are unit circle around origin, one can take a consistent solution to the above two equations (\ref{eq:saddleh},\ref{eq:saddlekl}) as
\be
z(x) = e^{i \q(x)}, \quad w(x) = e^{-i\q(x)}, \quad 
t=s=0
\ee
and
\be
k_n' = l_n'= {\b_n \r_n\over n}, \quad \text{where} \quad
Z_n = \int d\q \r(\q) \cos n\q \equiv\r_n.
\ee
Defining $\rho(z)=\frac{dx}{dz}$ equation (\ref{eq:saddlepoint}) can be written as
\ben\label{eq:saddlepoint2}
h(z) = \Xint{\ominus}_C dz' \rho(z') \frac{z}{z-z'} + \sum_n \frac{n k_n'}{Z_n}z^n.
\een
We define a quantity $\bar h(z)$
\ben\label{eq:hbar}
\bar h(z) = \oint_C dz'\rho(z') \frac{z}{z-z'} + \sum_n \frac{n k_n'}{Z_n}z^n.
\een
$\bar h(z)$ is analytic everywhere in complex $z$ plane except inside the support of $\rho(z)$ on $C$. It has a branch cuts along the support of $\rho(z)$. If $h_+(z)$ and $h_-(z)$ are values of $\bar h(z)$ on either sides of branch cuts then we have
\ben\label{eq:hpm}
h_{\pm}(z) = + \sum_n \frac{n k_n'}{Z_n}z^n + \oint_C dz'\rho(z') \frac{z}{z-z'} \pm i \pi z \rho(z).
\een
Considering contour $C$ to be unit circle we have
\ben\label{eq:hpmcircle}
\sum_n \frac{n k_n'}{Z_n}{e^{i n \theta}} + \dashint_{-\pi}^{\pi}d\theta \rho(\theta')\frac{e^{i\theta}}{e^{i\theta}-e^{i\theta'}} - h_{\pm}(\theta)\pm \pi \rho(\theta) =0
\een
Evaluating the partition function (\ref{eq:pftotal}) on equations (\ref{eq:saddlepoint},\ref{eq:saddleh}, \ref{eq:saddlekl}) we find,
\begin{eqnarray}\label{eq:zevbasis}
\begin{split}
Z &=  \int [\cD \q] \exp\lB N^{2}\sum_{n=1}^{\infty}
2\frac{\b_n}{n}\r_n
+N^{2}\frac{1}{2}\Xint-d\q\rho(\q)\Xint-d\q'
\rho(\q')\ln\left| 4\sin^{2}\lb\frac{\q-\q'}{2}\rb\right|\rB.
\end{split}
\end{eqnarray}
Thus we see that the partition function is exactly same as the partition function of one plaquette model written in eigenvalue basis (equation (\ref{eq:Zevbasis})) with $\rho(\theta)$ being the eigenvalue density. Hence, we identify the auxiliary variables we used to write the character of symmetric group with eigenvalues of the corresponding unitary matrix model involved.

Separating the saddlepoint equation (\ref{eq:hpmcircle}) into real and imaginary part we see that the imaginary part of this equation is same as the saddle point equation (\ref{eq:saddeq2}) for $\rho(\theta)$. Hence we are left with only the real part, which is given by,
\ben\label{eq:boundrel}
h_{\pm}(\theta) = \frac12 +\sum_n \beta_n \cos n \theta \pm \pi \rho(\theta).
\een
The above two relations can be written as a solution of a quadratic equation of $h$
\ben
h^2 -2S(\theta) h+S(\theta)-\pi^2 \rho^2(\theta)=0
\een
where $S(\theta)=\frac12 +\sum_n \beta_n \cos n\theta$.

\section{Spectral curve}
\label{app:speccurve}

Complexifying the above equation by replacing $\theta \ra -i \ln z$ yields,
\ben\label{eq:hzeqn}
h(z)^2 - 2 S(z) h(z) + S^2(z)-\pi^2 \rho^2(z) =0, \quad S(z)=  \frac12 + \sum_n \frac{\beta_n}2 \left(z+z^{-1}\right).
\een
$h(z)$ is defined on a two-sheeted cover of our original complex plane. The two sheets correspond to the two solutions of the quadratic equation (\ref{eq:hzeqn}). The solutions are given by
\ben
h_{\pm}(z) = S(z)\pm \pi \rho(z).
\een
The analytic properties of $h_{\pm}(z)$ on two Riemann sheets depends on the $\rho(z)$. $S(z)$ is a known meromorphic function with pole at $z=0$. For no-gap solution $\pi\rho(z)=S(z)$ and hence $h_+(z)=2S(z)$ and $h_(z)=0$. Therefore both $h_+(z)$ and $h_-(z)$ are analytic on both the Riemann sheets. There is no way one can connect these two Riemann sheets. The resulting geometry is given by $S^2 \cup S^2$ once we identify the points at infinity for both the Riemann sheets separately. 

For a $s$-gap solution $\rho(z)$ has the following form \cite{duttadutta}
\ben
\rho(z) = \frac{\sqrt{F(z)}}{2\pi}, \quad \where \ F(z)= f(z) \prod_{i=1}^s \left(z+a_i+\frac1z\right), 
\een
with $a_i\neq a_j$ for $i\neq j$. $a_i$ is also real and $-2\leq a_i\leq2, \ \forall i$. The function $f(z)$ is analytic. Hence $h_{\pm}(z)$ has $s$ branch cuts on unit circles on both the Riemann sheets. One can, therefore, glue these two Riemann sheets along $s$ cuts. As a result, the geometry is given by a genus $g=s-1$ Riemann surface.

\section{Asymmetric solutions}
\label{app:asymetricsol}

In large $k$ limit the matrix model (\ref{eq:cZ}) renders another class of solution \cite{duttagopakumar}. This solution is given by
\ben \label{eq:uhasym}
\begin{split}
u(h) &= \frac{2}{\pi} \cos^{-1} \lb \frac{h+\xi-1/2}{2\sqrt{\xi h}}\rb, \quad \for \quad p\leq h\leq q\\
& = 0 \quad \text{otherwise}
\end{split}
\een
where,
\be
\sqrt p = \sqrt{\xi} - \frac1{\sqrt2}, \quad \tand \quad \sqrt q = \sqrt{\xi} + \frac1{\sqrt2}
\ee
and fugacity $z$ (or $a$) is given by 
\be
z=\frac{4\xi^2}{4\xi -1}.
\ee
The solution is valid for $\xi>1/2$. The Young diagrams for this distribution is not symmetric under transposition. 

This is a valid solution in the context of GWW model. In case of GWW model the sum in equation (\ref{eq:PFcGWR}) was over the representations of unitary group $U(N)$ for which the maximum number of boxes in the first column of a Young diagram is $N$. Therefore, the symmetric representation fails to be a valid solution of GWW when the first column saturates this bound. As a result, GWW model undergoes a third order phase transition at $\xi=1/2$, known as Gross-Witten-Wadia phase transition.

However we are dealing with partition function (\ref{eq:GCPF}) where the sum is running over the representations of symmetric group. In this case there is no such restriction on $N$ (number of boxes in the first column). Thus we do not see any such phase transition here.

\bibliographystyle{hieeetr}
\bibliography{PlancherelNotes.bib}

\end{document}

%% file: gdefn.tex
\newcommand{\Tr}{\text{Tr}}

\newcommand{\seff}[1]{S_{\text{eff}}[{#1}]}

\newcommand{\ben}{\begin{eqnarray}\displaystyle}
\newcommand{\een}{\end{eqnarray}}

\newcommand{\be}{\begin{equation}}
\newcommand{\ee}{\end{equation}}

\newcommand{\bc}{\begin{center}}
\newcommand{\ec}{\end{center}}

\newcommand{\eesp}{\end{split}}
\newcommand{\bsp}{\begin{split}}

\newcommand{\Rmnum}[1]{\expandafter\@slowromancap\romannumeral #1@}

\renewcommand{\b}{\beta}		

\newcommand{\m}{\mu}								
\renewcommand{\o}{\omega}	
\newcommand{\q}{\theta}	
	
\renewcommand{\r}{\rho}		

\renewcommand{\t}{\tau}		
\newcommand{\x}{\xi}								

\newcommand{\I}{\infty}

\newcommand{\cA}{\mathcal{A}}

\newcommand{\cD}{\mathcal{D}}

\newcommand{\cH}{\mathcal{H}}

\newcommand{\cP}{\mathcal{P}}
\newcommand{\cQ}{\mathcal{Q}}

\newcommand{\cY}{\mathcal{Y}}
\newcommand{\cZ}{\mathcal{Z}}

\newcommand{\lna}[1]{\ln\left( #1 \right)}

\newcommand{\ra}{\rightarrow}

\newcommand{\lB}{\left [}
\newcommand{\rB}{\right ]}
\newcommand{\lb}{\left (}
\newcommand{\rb}{\right )}

\newcommand{\for}{\text{for}}
\newcommand{\where}{\text{where}}
\newcommand{\with}{\text{with}}
\newcommand{\tand}{\text{and}}

\newcommand{\bensp}{\begin{eqnarray}\begin{split}}
\newcommand{\eensp}{\end{eqnarray}\end{split}}

\newcommand{\bnm}{\begin{matrix}}
\newcommand{\enm}{\end{matrix}}

\def\Xint#1{\mathchoice
{\XXint\displaystyle\textstyle{#1}}%
{\XXint\textstyle\scriptstyle{#1}}%
{\XXint\scriptstyle\scriptscriptstyle{#1}}%
{\XXint\scriptscriptstyle\scriptscriptstyle{#1}}%
\!\int}
\def\XXint#1#2#3{{\setbox0=\hbox{$#1{#2#3}{\int}$ }
\vcenter{\hbox{$#2#3$ }}\kern-.6\wd0}}

\def\dashint{\Xint-}

\newcommand{\ket}[1]{|#1\big>}
\newcommand{\bra}[1]{\big<#1|}

\newcommand{\braket}[2]{\big<#1|#2\big>}

\newcommand{\dow}{\partial}

